\documentclass[twocolumn,english, preprint, onecolumn, superscriptaddress]{revtex4-1}
\usepackage[T1]{fontenc}
\usepackage[latin9]{inputenc}
\usepackage{xcolor}
\usepackage{babel}
\usepackage{array}
\usepackage{amsmath}
\usepackage{graphicx}
\PassOptionsToPackage{normalem}{ulem}
\usepackage{ulem}
\usepackage[unicode=true,pdfusetitle,
 bookmarks=true,bookmarksnumbered=false,bookmarksopen=false,
 breaklinks=false,pdfborder={0 0 1},backref=false,colorlinks=true]
 {hyperref}
\usepackage{breakurl}

\makeatletter

\providecommand{\tabularnewline}{\\}
\providecolor{lyxadded}{rgb}{0,0,1}
\providecolor{lyxdeleted}{rgb}{1,0,0}
\usepackage[caption=false]{subfig}
\usepackage{multirow}
\hypersetup{urlcolor = blue}
\hypersetup{citecolor = blue}
\hypersetup{linkcolor = blue}

\@ifundefined{showcaptionsetup}{}{%
 \PassOptionsToPackage{caption=false}{subfig}}
\usepackage{subfig}
\makeatother

\begin{document}

\title{Frequency Multiplexing for Deterministic Heralded Single-Photon Sources }

\author{Chaitali Joshi}

\affiliation{Department of Applied Physics and Applied Math, Columbia University,
New York City, USA }

\affiliation{Applied and Engineering Physics, Cornell University, Ithaca, NY,
USA}

\author{Alessandro Farsi}

\affiliation{Department of Applied Physics and Applied Math, Columbia University,
New York City, USA }

\author{St\'ephane Clemmen}

\affiliation{Laboratoire d\textquoteright Information Quantique, Université Libre
de Bruxelles, Bruxelles, Belgium}

\author{Sven Ramelow}

\affiliation{Institut für Physik, Humboldt-Universität zu Berlin, Berlin, Germany}

\author{Alexander L. Gaeta}
\email{Corresponding author: alg2207@columbia.edu}

\selectlanguage{english}%

\affiliation{Department of Applied Physics and Applied Math, Columbia University,
New York City, USA }
\begin{abstract}
Single-photon sources based on optical parametric processes have been
used extensively for quantum information applications due to their
flexibility, room-temperature operation and potential for photonic
integration. However, the intrinsically probabilistic nature of these
sources is a major limitation for realizing large-scale quantum networks.
Active feedforward switching of photons from multiple probabilistic
sources is a promising approach that can be used to build a deterministic
source. However, previous implementations of this approach that utilize
spatial and/or temporal multiplexing suffer from rapidly increasing
switching losses when scaled to a large number of modes. Here, we
break this limitation via frequency multiplexing in which the switching
losses remain fixed irrespective of the number of modes. We use the
third-order nonlinear process of Bragg scattering four-wave mixing
as an efficient ultra-low noise frequency switch and demonstrate multiplexing of three frequency modes.
We achieve a record generation rate of $4.6\times10^{4}$ multiplexed
photons per second with an ultra-low{\normalsize{} $g^{(2)}(0)=0.07$},
indicating high single-photon purity. Our scalable, all-fiber multiplexing
system has a total loss of just 1.3 dB independent of the number of
multiplexed modes, such that the 4.8 dB enhancement from multiplexing
three frequency modes markedly overcomes switching loss. Our approach
offers a highly promising path to creating a deterministic photon
source that can be integrated on a chip-based platform. 
\end{abstract}
\maketitle

\section{Introduction}

Deterministic and high quality single-photon sources are essential
to photonic quantum technologies including communications and information
processing. An ideal single-photon source should emit indistinguishable
photons in well-defined spatio-temporal and spectral modes with high
probability and negligible multi-photon noise. Efforts to build such
sources have focused primarily on the following two approaches: sources
that rely on nonlinear processes such as spontaneous parametric down
conversion (SPDC) or four-wave mixing, and single emitters such as
quantum dots, color centers and cavity-coupled atoms and ions \cite{eisaman_invited_2011-1}.
With recent engineering efforts for improved fabrication and control
of individual emitters, quantum dots with high brightness and photon
purity have been demonstrated \cite{loredo_scalable_2016-1,ding_-demand_2016,somaschi_near-optimal_2016-1}.
However, these sources require cryogenic cooling, lack spectral tunability
and are highly sensitive to the host solid-state environment, leading
to distinguishability between different emitters \cite{aharonovich_solid-state_2016}. 

Parametric sources on the other hand can be easily adapted to a wide
variety of experimental conditions and have been used for pioneering
quantum information experiments including quantum teleportation, loop-hole-free
Bell tests and boson sampling \cite{ma_quantum_2012,giustina_significant-loophole-free_2015,broome_photonic_2013,spring_boson_2013,shalm_strong_2015}.
These sources operate at room temperature and provide highly indistinguishable
photons with flexible control over the spectral and temporal properties
of the photons \cite{kwiat_new_1995,kwiat_ultrabright_1999,tanzilli_highly_2001,fedrizzi_wavelength-tunable_2007}.
Such sources have proved to be highly versatile, producing photons
spanning the visible to the infrared, with bandwidths ranging from
a few hundred kHz to a few THz \cite{rambach_sub-megahertz_2016,nasr_ultrabroadband_2008,odonnell_observation_2007}.
Moreover, parametric sources can be fully integrated onto monolithic
CMOS-compatible platforms to generate narrow band entangled photons
with high brightness \cite{reimer_integrated_2014,reimer_generation_2016,ramelow_silicon-nitride_2015-1}.
However, these sources are fundamentally limited by multi-photon generation,
resulting in probabilistic operation with a maximum heralding efficiency
of 25\% from a single source.

Active feed-forward switching of photons from multiple identical sources
is a promising technique that can overcome the probabilistic operation
of a single source \cite{migdall_tailoring_2002,jeffrey_towards_2004,pittman_single_2002,shapiro_-demand_2007,mower_efficient_2011}.
By operating individual sources in a regime with low pair production
probability, such schemes allow for increasing the single-photon probability
without additional multi-photon generation. A key requirement for
efficient multiplexing is a low-loss $N\times1$ switching network
that accommodates a sufficiently large number of modes $N$ to achieve
deterministic operation. Deterministic operation can be achieved with
as few as $N=17$ multiplexed modes with a lossless switching network
and photon-number resolving (PNR) detectors \cite{christ_limits_2012-2}.
Recently, there have been a number of promising demonstrations of
multiplexed sources using the spatial and temporal degrees of freedom
of a photon \cite{ma_experimental_2011,broome_reducing_2011,collins_integrated_2013,mendoza_active_2016,xiong_active_2016,kaneda_time-multiplexed_2015-1,francis-jones_all-fiber_2016}.
However, for both spatial and temporal multiplexing, switching losses
increase with the number of modes $N$, which deteriorates enhancement
achieved from multiplexing beyond a few modes. Deterministic operation
is therefore challenging to achieve without the use of bulky free-space
setups \cite{kaneda_time-multiplexed_2015-1}. 

Here, we propose and demonstrate an alternative scheme using frequency
multiplexing where losses do not scale with the number of modes. Frequency
multiplexing allows for multiple switching operations in a single
spatial mode, thus effectively implementing an $N\times1$ switch
in a monolithic optical structure such as a single mode fiber or waveguide.
Therefore, distinct from other schemes, switching losses remain fixed
irrespective of the number of multiplexed modes $N$. We use tunable
quantum frequency translation via Bragg scattering four wave mixing
(BS-FWM) \cite{inoue_tunable_1994,mckinstrie_translation_2005,mcguinness_quantum_2010,li_efficient_2016},
which we realize with close to unity efficiency and ultra-low noise
\cite{farsi_low-noise_2015,clemmen_ramsey_2016}. Tunable frequency
translation allows us to perform active ``frequency switching''
of multiple frequency channels. We have previously proposed frequency
multiplexing and reported preliminary experimental results using two
frequency modes \cite{joshi_frequency_2016}. Here, we present a complete
characterization of our frequency multiplexing scheme, including theoretical
analysis of the scaling performance for large $N$. We present a proof-of-principle
demonstration of frequency multiplexing using three frequency modes
in an entirely fiber-based setup that leverages on low-loss off the
shelf dense wavelength division multiplexing (DWDM) components. With
this low-loss and low-noise setup we achieve record generation rates
of 46 kHz multiplexed photons with coincidences-to-accidentals ratio
exceeding 100 and $g^{(2)}(0)$ of 0.07. BS-FWM is efficiently tunable
over a large bandwidth of more than 1 THz and therefore our system
can be scaled to include a large number of frequency modes, which
is required for deterministic photon generation using multiplexing. 

We note that recently, multiple research groups have proposed the
use of frequency multiplexing as a resource for both continuous variable
\cite{roslund_wavelength-multiplexed_2014,humphreys_continuous-variable_2014}
and circuit-based single-photon QIP applications \cite{sinclair_spectral_2014,lukens_frequency-encoded_2017}.
These proposals emphasize the strong potential of frequency multiplexing
for addressing the scaling losses and resource overheads in quantum
systems. However, most proposals rely on electro-optic modulators
(EOMs) to frequency translate single photons. Recent work \cite{puigibert_heralded_2017}
discusses a spectrally multiplexed single-photon source using EOMs,
but no enhancement in the single-photon rate is demonstrated due to
high system losses. In addition, EOMs typically have a limited time-bandwidth
product close to unity, limiting the maximum frequency shift and the
bandwidth of the target pulses. This significantly limits practical
implementations to a few frequency modes while exacerbating photon
loss due to narrow filtering. Alternatively, our implementation of
BS-FWM allows for tunable conversion over 1 THz with an acceptance
bandwidth of 100 GHz with few nanosecond pump pulses, which addresses
these issues.  

\section{Principle and Theory: }

\begin{figure*}[p]
\begin{centering}
\includegraphics[width=0.8\paperwidth]{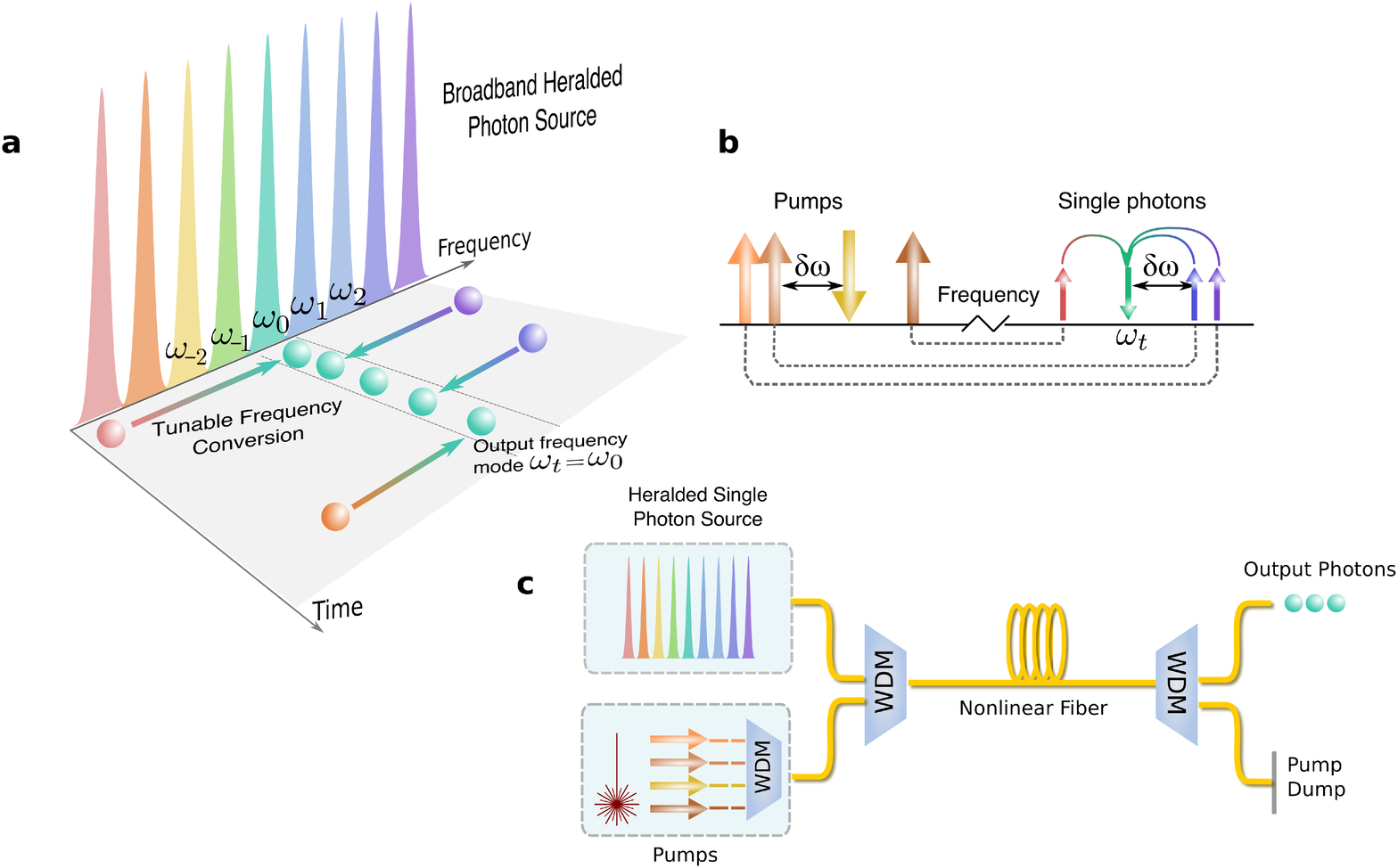}
\par\end{centering}
\caption{\label{fig:Principle-of-frequency}\textbf{Principle of a frequency
multiplexed single-photon source.} \textbf{a)} Multiple narrowband
frequency channels $\left\{ \omega_{0},\omega_{1}...\right\} $ are
extracted from a broadband single-photon source. Tunable frequency
conversion is used to convert photons from different channels to a
common target frequency mode $\omega_{t}$. \textbf{b)} Tunable frequency
conversion using Bragg scattering four-wave mixing (BS-FWM): Two strong
classical pumps drive the interaction between the input and target
$(\omega_{t})$ single-photon fields. The frequency separation $\delta\omega$
between the pump fields determines the frequency shift of the single
photons, and additional pump fields can be used to increase the number
of possible values of $\delta\omega$. \textbf{c)} Fixed-loss operation
of frequency multiplexing: single photons and BS-FWM pumps are combined
using wavelength division multiplexers (WDM), and all active frequency
switching takes place in a single nonlinear fiber/waveguide. Additional
channels can be added by introducing additional pumps, without introducing
losses in the path of the single photons. }

\end{figure*}

Figure \ref{fig:Principle-of-frequency} illustrates our frequency
multiplexing scheme. A single source that generates broadband frequency
correlated photon pairs is used to create narrowband frequency channels
$\left\{ \omega_{0},\omega_{1}...\right\} $. One photon from the
pair (heralding photon, not shown) is used to herald the presence
of the signal photon. Due to energy conservation, the two photons
are correlated in frequency, with the heralding photon providing information
about the frequency of the signal photon. This heralding information
is used to translate the frequency of the signal photon to the target
frequency channel $\omega_{t}$ using tunable frequency conversion.
We thus effectively implement an active frequency switch to route
photons from multiple frequency bins to a single output frequency
channel. In order to be viable as an $N\times1$ switch for large
$N$, the tunable frequency conversion must be efficient over a sufficiently
large bandwidth. For this purpose, we use BS-FWM, a third-order nonlinear
parametric process involving the coherent interaction between two
quantum fields at different frequencies mediated by two strong classical
pumps \cite{mcguinness_quantum_2010}. Contrary to frequency conversion
based on parametric amplification, BS-FWM is theoretically noiseless
and preserves all quantum properties of the translated photons. BS-FWM
allows for independent control of the input and target frequencies
by selectively activating auxiliary pumps in the interaction (see
Figure \ref{fig:Principle-of-frequency}b). Since phase matching can
be achieved by symmetric placement of the classical pumps and quantum
fields about the zero-dispersion wavelength of the nonlinear medium,
the same setup can be reconfigured to target different frequency shifts
by tuning the pump wavelength (see supplementary section I). For efficient
conversion, it is critical that the bandwidth of individual channels
be less than the acceptance bandwidth $\Delta\nu_{BS}$ of the BS-FWM
process for two fixed pumps. This all-optical frequency switch can
support ultrafast operation, with the repetition rate limited only
to the inverse bandwidth $1/\Delta\nu_{BS}.$ Finally, we note that
all frequency switching takes place in a single spatial mode (nonlinear
fiber/waveguide), as shown in Figure \ref{fig:Principle-of-frequency}c.
As additional channels only require additional BS-FWM pumps, no scaling
losses are introduced in the path of the single photons.  

To understand clearly the characteristics of our frequency multiplexing
scheme, we analyze how the performance scales with increasing $N$.
Several architectures have been explored for active $N\times1$ switching
of photons in spatial and temporal multiplexing schemes. Typically,
these architectures use $2\times2$ switches as building blocks for
a general $N\times1$ switch. We compare the performance of the fixed
loss scheme with the log-tree network which is generally used for
spatial multiplexing and multi-pass binary switches (or storage cavities)
generally used in temporal multiplexing \cite{xiong_active_2016,kaneda_time-multiplexed_2015-1}.
An $N\times1$ log-tree network has a depth $\lceil\log_{2}N\rceil$.
Assuming a switching efficiency of $\eta_{s}$ per switch, the losses
scale as $\eta_{s}^{\lceil\log_{2}N\rceil}$\cite{bonneau_effect_2015}.
The losses from multi-pass binary switching scale exponentially as
$\eta_{s}^{N}$ in the worst case, but we consider an optimized implementation
as in Ref. \cite{kaneda_time-multiplexed_2015-1}. For our fixed-loss
scheme, the switching losses are $\eta_{s}$ irrespective of $N$.

\begin{figure}[t]
\subfloat[]{\centering{}\includegraphics[width=0.5\textwidth]{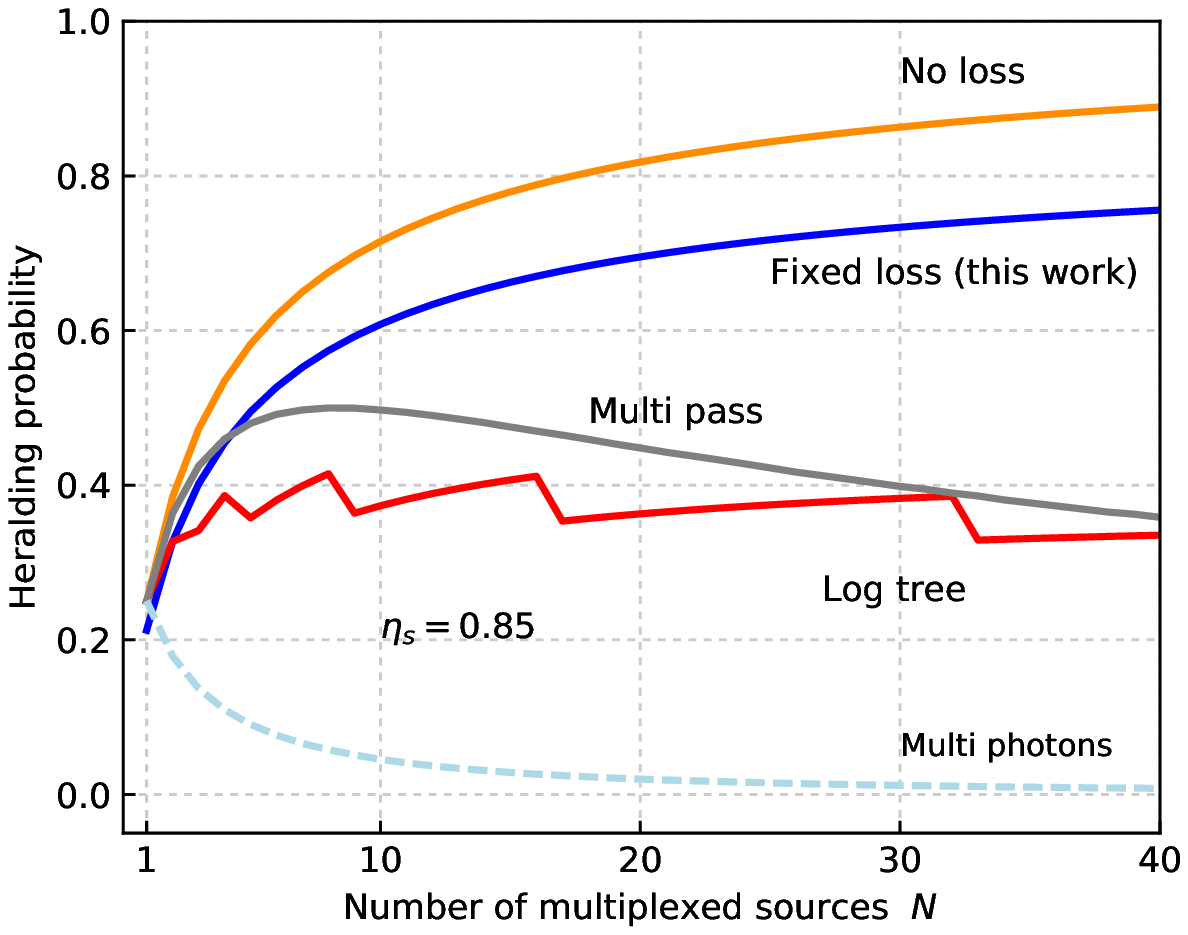}}\subfloat[]{\centering{}\includegraphics[width=0.5\textwidth]{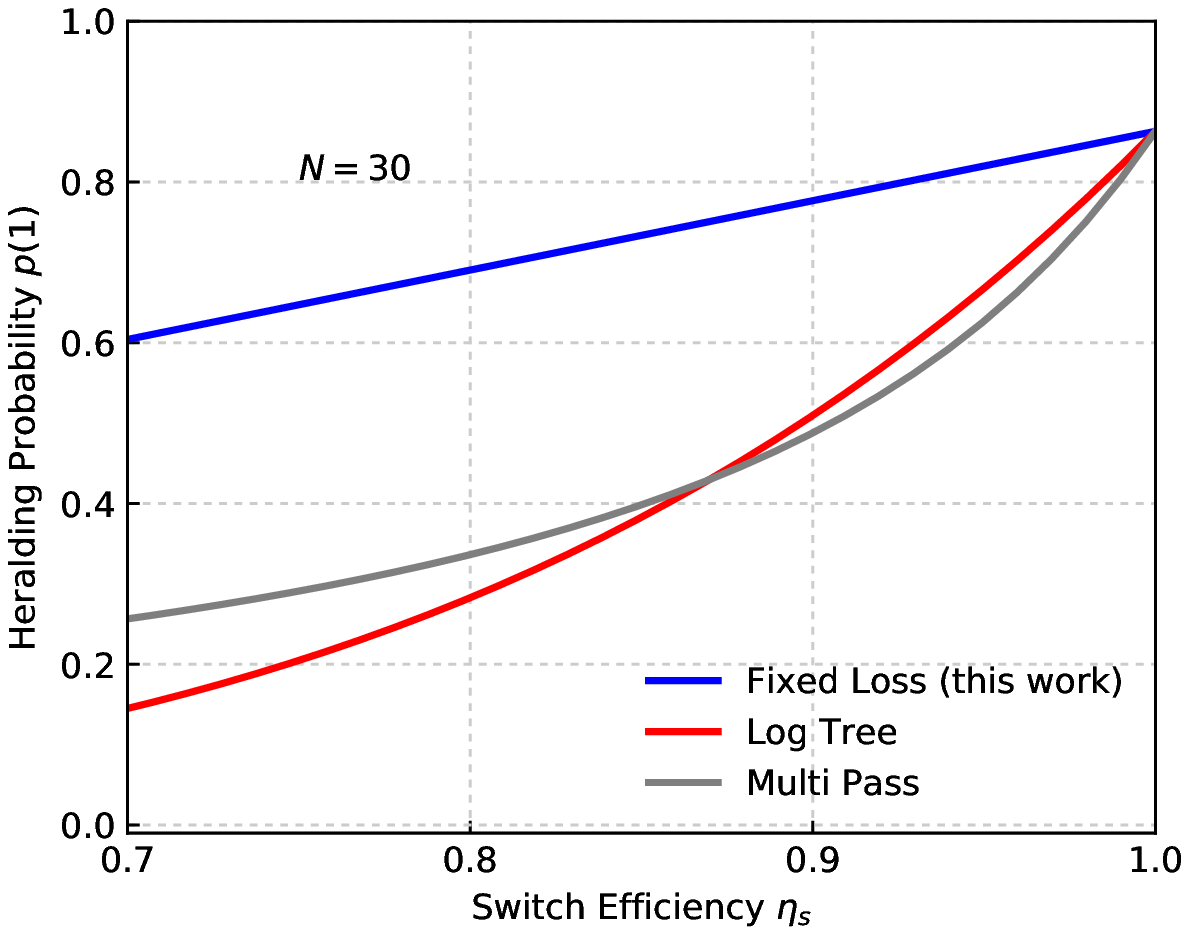}}

\caption{\label{fig:Theory_results}\textbf{Theoretical prediction of scaling
performance for various switching schemes}. \textbf{a)} The maximum
single-photon heralding probability for a single source ($N=1$) is
0.25. For an efficiency of $\eta_{s}=$ 0.85 (0.7 dB loss) per switch,
the single-photon emission probability for both log-tree and multi-pass
schemes reaches a maximum of $0.41,0.50$, respectively, and then
saturates for large $N$ due to scaling losses. In contrast, additional
multiplexed sources always result in improved performance for the
fixed-loss scheme, with $p_{mux}(n=1)=0.75$ for $N=40$ sources.
The multi-photon emission probability $p_{mux}(n>1)$, ignoring switching
losses, is shown as the dashed light blue curve, and is less than
$1\%$ for $N=40$ modes. \textbf{b)} Single-photon heralding probability
for various switching schemes as a function of switch loss, for a
fixed number of sources $N=30$. The fixed-loss scheme is significantly
more robust to variability in switching losses. Note that the maximum
heralding probability for $\eta_{s}=1$ is not equal to 1 since the
heralding detectors are non-photon-number resolving. }
\end{figure}

Figure \ref{fig:Theory_results}a shows the scaling performance of
various schemes. We assume a switching efficiency $\eta_{s}=0.85$
($0.7$ dB loss) per switch and all other components, including detectors,
are assumed to be ideal. We optimize the emission probability per
source $p_{single}(n=1)$ for each $N$ (see supplementary information
section II). The maximum heralding probability for a single source
($N=1$) is $0.25$. For both log-tree and multi-pass schemes, the
single photon probability reaches a maximum of $0.41$ and $0.50$,
respectively, and saturates due to switching losses for less than
$N=10$ multiplexed sources. In contrast, for the fixed-loss scheme,
additional multiplexed sources always result in an improvement in
the single-photon heralding probability, with a heralding probability
of 0.60 for $N$= 10 sources. For $N$= 40 sources, the fixed-loss
scheme achieves $p_{mux}(n=1)=$ 0.75, compared with a maximum of
$0.89$ with a no-loss ideal switching network. Our scheme therefore
has an advantage in the intermediate regime of 10 to 20 multiplexed
modes as well as asymptotically for large $N$. In order to quantify
the effects of practical variability in switching efficiency in implementations
of multiplexed sources, we analyze the sensitivity of different schemes
to switching losses in Figure \ref{fig:Theory_results}b. For a moderate
increase in losses to $1.2$ dB per switch and 30 multiplexed modes
$(\eta_{s}=0.75,N=30)$, the single photon probability drops significantly
from $0.86$ ($\eta_{s}=1$) to $0.21,0.29$ for the log-tree and
multi-pass schemes, but is reduced only moderately to $0.65$ for
the fixed-loss scheme. Thus, the frequency multiplexing scheme is
significantly more robust to switching losses as compared to competing
switching architectures in other multiplexed sources. 

\section{Experimental setup}

\begin{figure*}[t]
\begin{centering}
\includegraphics[width=1\textwidth]{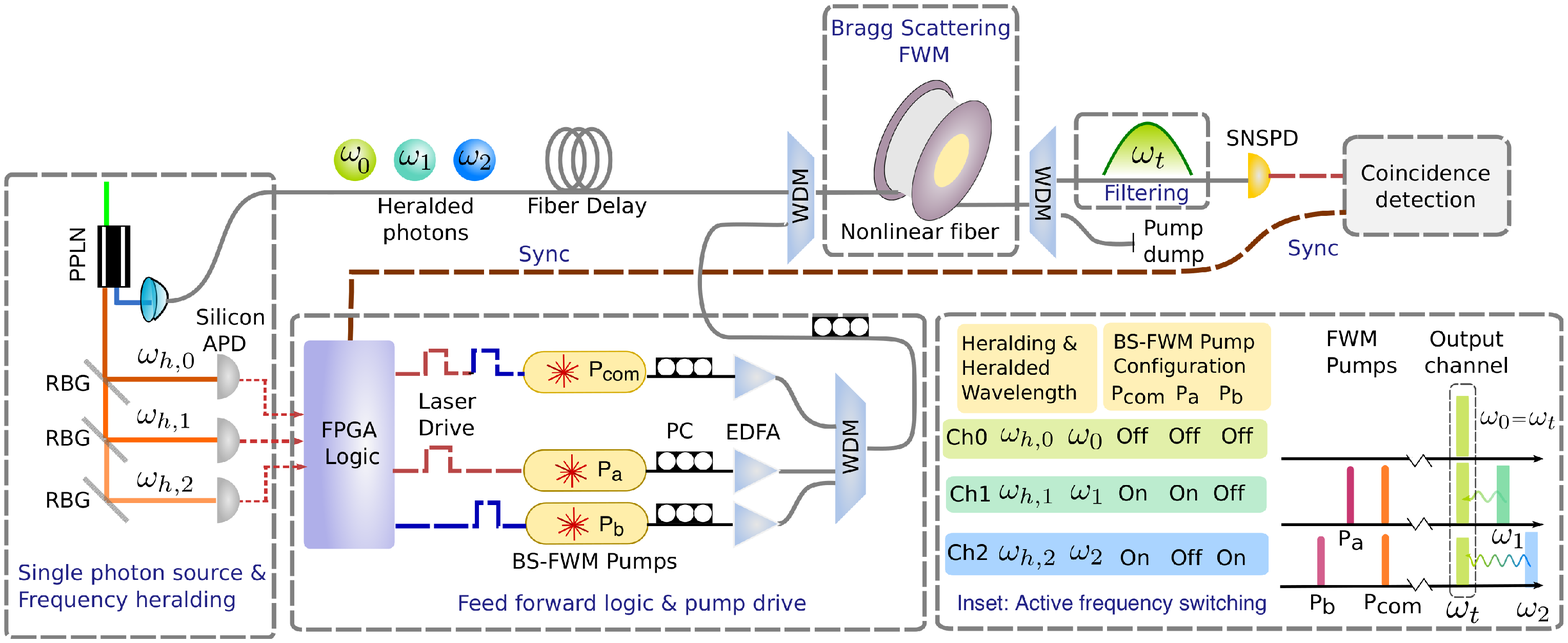}
\par\end{centering}
\caption{\label{fig:Experimental-setup}\textbf{Experimental setup for multiplexing
of three frequency modes}. PPLN - periodically poled lithium niobate,
RBG - reflecting Bragg grating , FPGA - Field programmable gate array,
PC: polarization control, EDFA: erbium-doped fiber amplifier, WDM-
wavelength division multiplexer, BS-FWM - Bragg scattering four-wave
mixing, SNSPD: superconducting nanowire single-photon detectors. A
PPLN crystal is pumped with a CW laser at 543 nm, generating photon
pairs at 940 nm (heralding photons) and 1280 nm (heralded photons).
The heralding photons are filtered into three channels \{$\omega_{h,0},$$\omega_{h,1}$,
$\omega_{h,2}$\} with 100 GHz bandwidth, with corresponding heralded
signal photons as \{$\omega_{0},\omega_{1},\omega_{2}$\}. An FPGA
is used to process this heralding information and selectively activate
the BS-FWM pumps (see inset), such that the heralded photons are switched
to the target frequency $\omega_{t}$. The signal photons are combined
with the BS-FWM pumps using WDMs, and sent to the nonlinear fiber.
A free-space filtering setup extracts photons at $\omega_{t}$ at
the output, which are then sent to an SNSPD. A time-tagging module
is used for coincidence measurements between the FPGA processed heralding
trigger (sync) and the output of the SNSPD. }
\end{figure*}

We experimentally demonstrate multiplexing of three frequency modes.
Figure \ref{fig:Experimental-setup} shows our experimental setup.
Our multiplexed source is based on broadband SPDC in a periodically-poled
lithium-niobate crystal (PPLN) pumped with a 543-nm CW-laser, generating
photon pairs at 940 nm (heralding photons) and 1280 nm (heralded signal
photons). The heralding photons are sent to a filtering setup consisting
of reflecting Bragg gratings (RBG), creating three channels $\text{CH0, \text{CH1, CH2}}$
with heralding photons at $\omega_{h,0},$ $\omega_{h,1}$, $\omega_{h,2}$,
respectively, with 100-GHz bandwidth (see supplementary section III).
Each channel is collected into a single-mode fiber and sent to a silicon
avalanche single-photon detector, which provides heralding information
to the logic circuit. The source crystal temperature is tuned to maximize
photon pair production at $\omega_{1}=1280.65$ nm and $\omega_{2}=1280.1$
nm, and the pair production at $\omega_{0}=1284.45$ nm is lower by
a factor of 0.65. The heralded signal photons \{$\omega_{0}=1284.45\text{ {nm}},\omega_{1}=1280.65\text{ {nm}},\omega_{2}=1280.1\text{ {nm}}$\}
are injected into the multiplexing setup, comprised of a 100-m nonlinear
fiber, wavelength-division-multiplexing couplers and a pump filter
(total losses 2.2 dB). A single channel centered at $\omega_{t}=1284.45$
nm and 100 GHz wide, is selected with a tunable grating and then sent
to a superconducting nanowire single-photon detector (SNSPD) with
a quantum efficiency of 53\%. 

The nonlinear process of BS-FWM is driven by two pump waves generated
by distributed feedback lasers diodes, which determine the frequency
shift and hence the input and output frequency channels. The diodes
are driven with a 5-ns-long pulsed current source, and the optical
pulses (for convenience aligned to the C-band ITU grid) are amplified
to a peak level of 10 W via cascaded erbium-doped fiber amplifiers
(EDFA). The pump pulses are combined together, temporally synchronized
and aligned in polarization. In order to achieve fast switching operations,
we utilize lasers at predetermined wavelengths that are selectively
turned on and off via a fast logic circuit controlled by a field programmable
gate array (FPGA) (see inset in Figure \ref{fig:Experimental-setup}).
We measure the conversion efficiency for both process $\omega_{1}\rightarrow\omega_{t}$
and $\omega_{2}\rightarrow\omega_{t}$ to be 93\% (see supplementary
section I).

\section{Results}

\begin{table}[t]
\begin{tabular}{|>{\centering}m{40pt}|>{\centering}p{40pt}|>{\centering}p{65pt}|>{\centering}p{65pt}|>{\centering}p{80pt}|>{\centering}p{80pt}|}
\hline 
 & \#Modes N & Multiplexing enhancement & Heralding Efficiency & Single-photon generation rate & BS-FWM repetition rate\tabularnewline
\hline 
a  & 3 & 2.2 & 4.6\% & 46 kHz & 1 MHz\tabularnewline
\hline 
b & 10 & 8.5 & 50\% & 2.5 MHz & 5 MHz\tabularnewline
\hline 
\end{tabular}

\caption{\label{tab:Performance-scaling}Performance of the frequency multiplexed
source \textbf{a)} current system with 3 modes and multiplexing system
loss of 1.3 dB ($\eta_{s}=0.75$) \textbf{b)} scaled source with 10
modes and improved system efficiency ($\eta_{s}=0.85,$ fiber-collection
and detection efficiency of 90\%)}
\end{table}

\subsection*{Heralded single-photon rate}

We first characterize the heralded single-photon rates as functions
of SPDC pump power for each individual channel and for the multiplexed
source, as shown in Figure \ref{fig:Exptresults}a. The multiplexed
(MUX) source has an enhanced coincidence rate by $4.8$ dB as compared
to the mean photon rate of the individual channels. This enhancement
significantly overcomes the losses of the setup (1.3 dB), resulting
in a net enhancement of $3.5$ dB $(220%\%)
$\%) in the heralded single-photon rate. At maximum SPDC pump power
(25 mW), we measure a heralding rate of 1 MHz with a brightness of
23 kHz detected coincidences per second. Supplementary section IV
provides detailed characterization of the system efficiency and losses.
We estimate a single-photon generation rate of 46 kHz after correcting
for detector efficiency $(3$ dB), which is the highest reported rate
for multiplexed photon sources to date. We note that although simply
increasing the pump power of the SPDC source can increase the single-photon
generation rate of a single source, this would lead to increased multi-photon
generation. 

\subsection*{Coincidences-to-accidentals ratio (CAR)}

We measure the coincidences to accidentals ratio (CAR), a standard
figure of merit to characterize the multi-photon generation of parametric
sources. Figure \ref{fig:Exptresults}b compares the CAR for the multiplexed
source and each individual channel. For fair comparison we also measure
the coincidence rate and CAR at $\omega_{t}$, directly from the SPDC
source, without the multiplexing setup in place (referred to as the
NoMUX source). We operate in a regime in which the single-photon count
rate is much higher than the dark-count rate of the detectors, and
therefore the accidental counts are dominated by multi-photon generation,
which is inversely proportional to the SPDC pump power. The multiplexed
source has a CAR that is a factor of 2 higher throughout as compared
to the NoMUX source. For low count rates, the multiplexed source has
a CAR exceeding $1000$ and remains high at $100$ at the maximum
count rate. These measurements confirm that the strong classical pumps
used in BS-FWM do not introduce significant spurious noise photons
even at a high pump trigger rate of 1 MHz. 

\subsection*{Single-photon purity ($g^{(2)}(0)$)}

\begin{figure}[t]
\vspace{-60pt}
\hspace{-60pt}
\begin{centering}
\subfloat[]{\centering{}\includegraphics[width=0.55\textwidth]{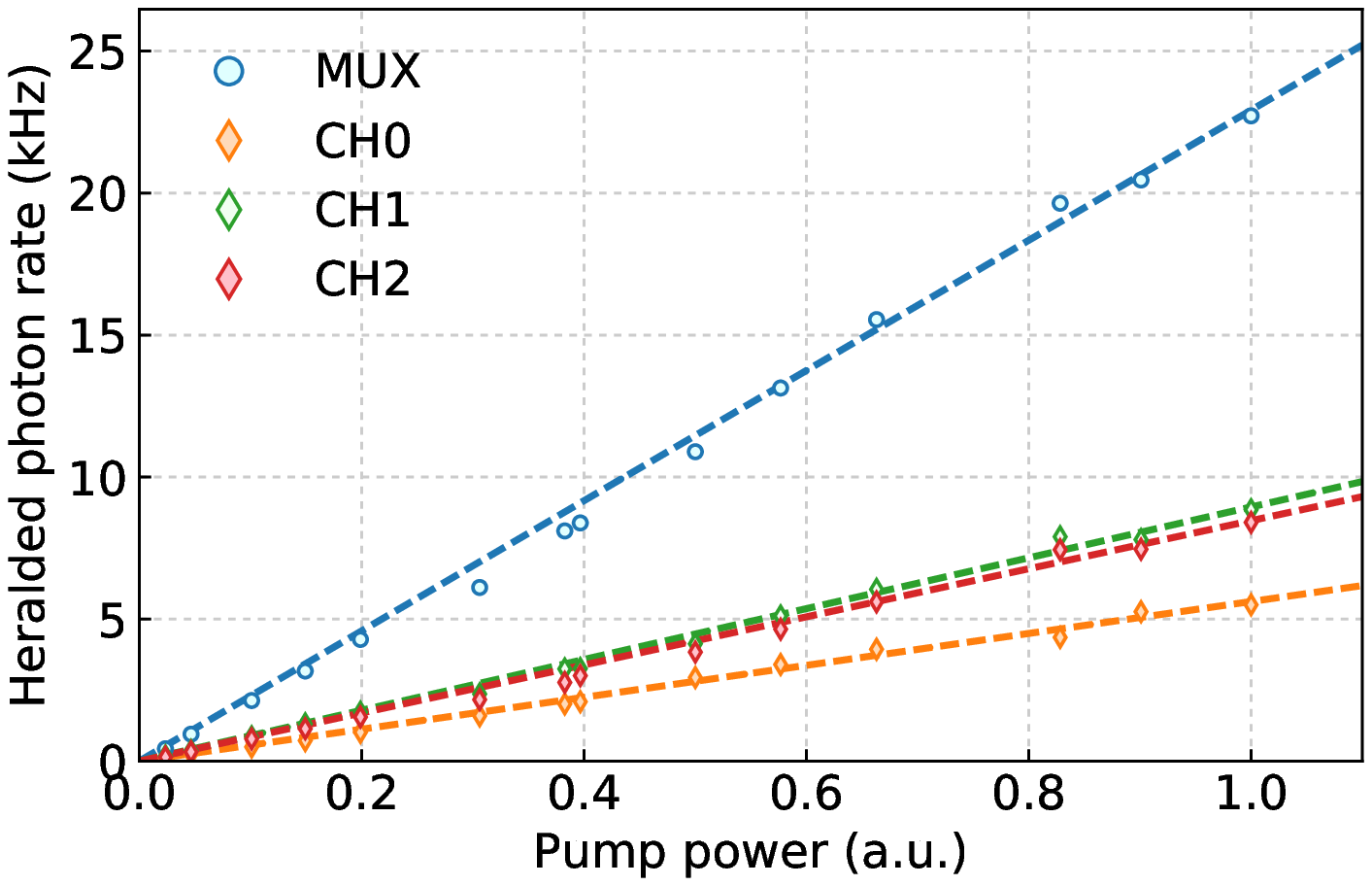}}\subfloat[]{
\begin{centering}
\includegraphics[width=0.55\textwidth]{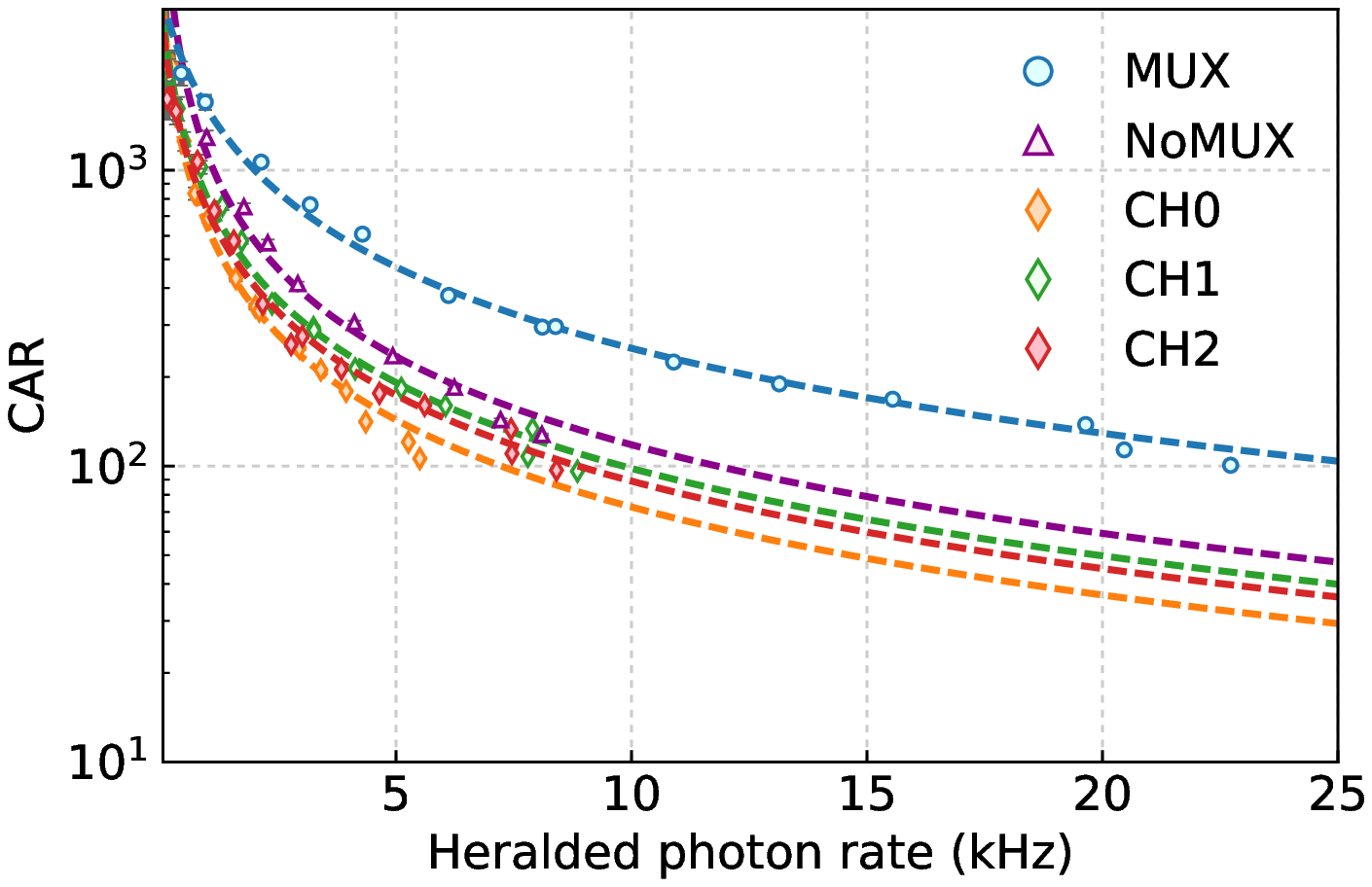}
\par\end{centering}
}\hfill{}\subfloat[]{
\centering{}\includegraphics[width=0.55\textwidth]{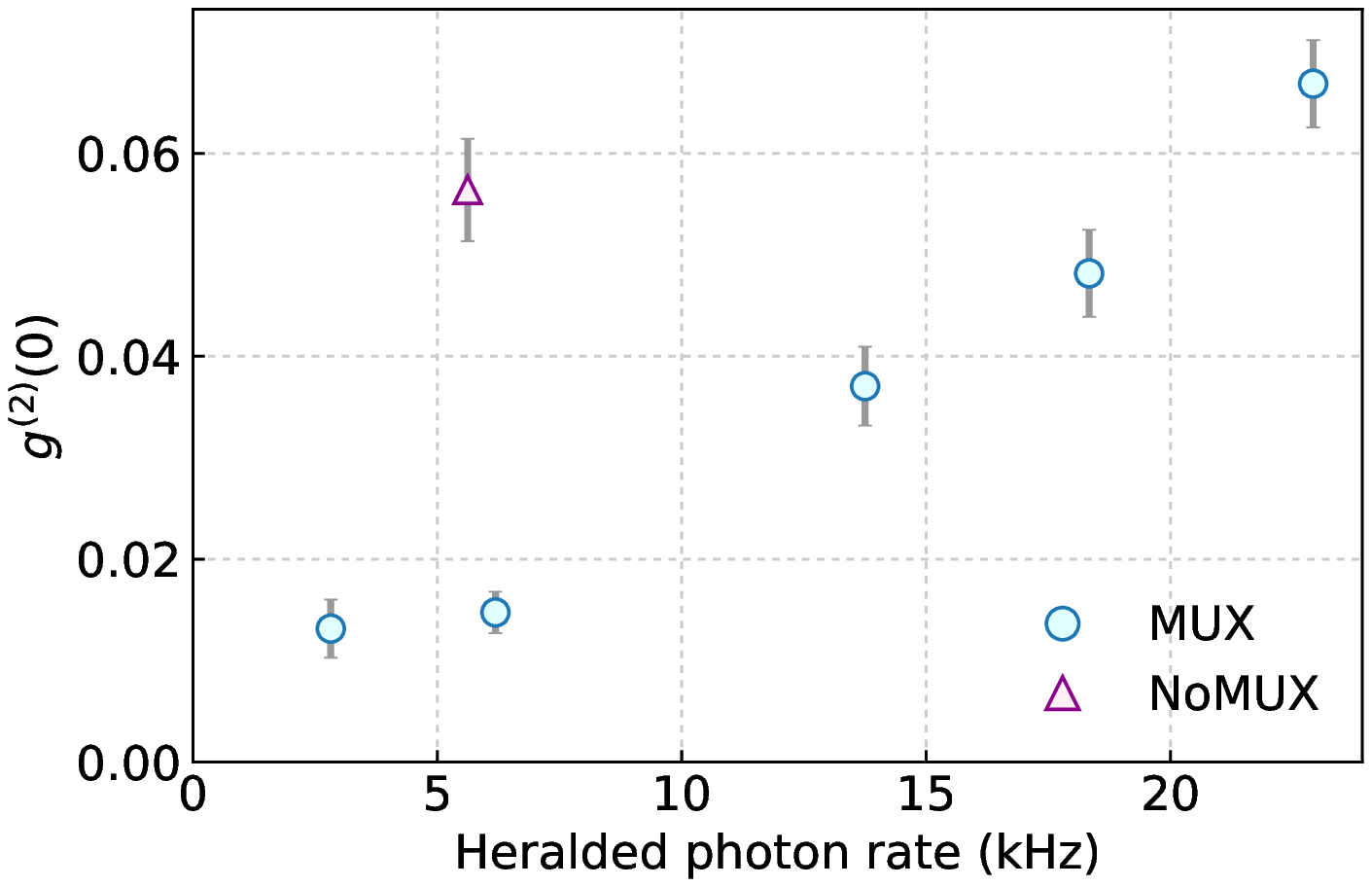}}
\par\end{centering}
\caption{\label{fig:Exptresults}\textbf{Experimental results from multiplexing
of three frequency modes.} Blue circles: multiplexed (MUX) source,
purple triangles: NoMUX source, diamonds: contributions from the individual
channels (orange - $\text{{CH0}}$, green - $\text{{CH1}}$, red -
$\text{{CH2}}$).\textbf{ a)} Total coincidence counts (heralded photons)
as a function of SPDC pump power: The multiplexed source has an enhanced
coincidence rate by $4.8$ dB as compared to the mean of the individual
channels and overcomes the losses of the setup ($1.3$ dB), with a
net enhancement of 220\%. At maximum SPDC pump power (25 mW), we measure
a heralded photon rate of 23 kHz from the multiplexed source.\textbf{
b) }Coincidences to accidentals ratio (CAR) vs coincidences: For a
fixed coincidence rate, the multiplexed source has a CAR that is a
factor of 2 higher as compared to the NoMUX source. For low coincidence
rates, the multiplexed source has a CAR exceeding 1000 and remains
high at 100 for large coincidence rates. \textbf{c)} Measurement of
$g^{(2)}(0)$: The multiplexed source has a low $g^{(2)}(0)$ of $0.07\pm0.005$
for large coincidence rates. For the same coincidence rate of 2.5
kHz the multiplexed source has an improved single-photon purity with
$g^{(2)}(0)$ of $0.015\pm0.002$ as compared to the NoMUX source
with a $g^{(2)}(0)$ of $0.056\pm0.005$. Error bars are estimated
using Poisson statistics. }
\end{figure}

Finally, we measure the purity of the photons from the multiplexed
source by the second-order correlation function $g^{(2)}=\left\langle \hat{N}_{a}\hat{N}_{b}\right\rangle /\left\langle \hat{N}_{a}\right\rangle \left\langle \hat{N}_{b}\right\rangle $
where $\hat{N_{a}}$ and $\hat{N_{b}}$ are photon number operators
corresponding to the two arms of a Hanbury-Brown-Twiss setup \cite{brown_correlation_1956}.
Figure \ref{fig:Exptresults}c shows the measured $g^{(2)}(0)$ for
the multiplexed source and the NoMUX source, for various heralded
photon rates. At the maximum heralded photon rate, the multiplexed
source has a low $g^{(2)}(0)$ of $0.07\pm0.005$. For the same low
heralded photon rate of 2.5 kHz, the multiplexed source and the NoMUX
source have $g^{(2)}(0)$ of $0.015\pm0.002$ and $0.056\pm0.005$
respectively. The average SPDC pump power required to achieve the
same photon rate is a factor of 3 lower for the multiplexed source
as compared to the NoMUX source, and therefore has significantly reduced
multi-photon generation. The improved single photon purity of the
multiplexed source is therefore a strong indicator of successful multiplexing. 

The performance of our frequency multiplexed source is comparable
with the best multiplexed source demonstrated to-date which implements
temporal multiplexing on a free-space optics platform\cite{kaneda_time-multiplexed_2015-1}.
A complete comparison with other relevant works has been included
in the supplementary section V. We achieve a record single-photon
generation rate of 46 kHz with an ultra-low $g^{(2)}(0)$ of 0.07,
compared with previously demonstrated 19.3 kHz with a high $g^{(2)}(0)$
of 0.48 \cite{kaneda_time-multiplexed_2015-1}. Due to the low loss
of our frequency switch (1.3 dB), we achieve a multiplexing enhancement
factor of 2.2 with just three frequency modes. We measure a raw heralding
efficiency of 2.3\% and detector-corrected efficiency of 4.6\%, which
is the highest amongst fiber-based and integrated multiplexed systems
demonstrated so far. This efficiency is mainly limited by the fiber-collection
and spectral filtering loss at the SPDC source and is independent
of our ``frequency switching'' setup. Collection efficiencies as
high as 90\% can be achieved by minimizing all transmission and filtering
losses, and careful mode-matching, which would correspond to an order
of magnitude improvement in the heralded single-photon rates \cite{ramelow_highly_2013,shalm_strong_2015}.
Another important figure of merit for comparing the different multiplexing
implementations is the maximum possible switching speed. In principle,
our all-optical frequency switch allows for efficient conversion with
repetition rates as high as the inverse of the BS-FWM acceptance bandwidth
(100 GHz in this system). Our current implementation can support a
repetition rate of 5 MHz and is only limited by the amplification
required for the BS-FWM pumps. This amplification requirement can
be reduced by increasing the BS-FWM interaction length or using highly
nonlinear fibers as the interaction medium, enabling significantly
higher repetition rates. 

\subsection*{Pulsed operation and scaling}

In order to obtain photons in well-defined temporal modes, pulsed
operation is necessary. The efficiency of our ``frequency switch''
is partially limited to $93\%$ due to the fluctuations in BS-FWM
pump power induced by the randomized trigger arising from CW operation
of the single-photon source. We measure efficiencies as high as $97\%$
using the same setup with periodic pump triggering. In our existing
fiber-based setup, we can incorporate up to 10 frequency modes without
a decrease in frequency conversion efficiency. We summarize the scaling
performance of our system in Table \ref{tab:Performance-scaling}.
With just 10 multiplexed modes, our system is capable of achieving
a single-photon heralding probability exceeding 50\% (per input pump
pulse) with a single-photon generation rate of 2.5 MHz (see supplementary
section VI). Finally, we note that using cavity-based sources with
the spectral line-width of the pump pulse matched to the cavity line-width,
it is possible to generate discrete uncorrelated joint spectral amplitudes
\cite{helt_spontaneous_2010}. Our system is therefore capable of
approaching the regime of deterministic photon generation in pure
spectral and temporal modes. 

\section{Discussion}

We have demonstrated a novel frequency multiplexed source with three
modes, using highly-efficient low-noise quantum frequency translation.
We emphasize that adding additional channels adds complexity only
to the BS-FWM pump configuration, and no new components need to be
added in the path of the single photons. This ensures that losses
remain independent of the number of multiplexed modes. The single
spatial mode operation of frequency switching maintains relative polarization
stability of photons from different channels from generation to detection,
ensuring that the photons are rendered indistinguishable after frequency
translation. BS-FWM is fully compatible with the existing optical
telecommunication architecture that harnesses dense wavelength division
multiplexing (DWDM). The applications of such low-loss high repetition
rate frequency multiplexing go beyond single-photon sources and can
prove to be highly advantageous for all-photonic quantum repeaters
that rely on active feed-forward heralding signals \cite{azuma_all-photonic_2015}.
Our scheme is also entirely adaptable to CMOS-compatible integrated
platforms. In particular, integrated comb sources where photons are
already confined in well-defined frequency bins can eliminate the
need for filtering \cite{reimer_integrated_2014,reimer_generation_2016,ramelow_silicon-nitride_2015-1}
while generating spectrally pure photons \cite{helt_spontaneous_2010}.
In addition, implementations of BS-FWM in nanophotonic waveguides
can significantly reduce pump power and amplification requirements
\cite{agha_low-noise_2012}. Frequency multiplexing can thus uniquely
harness both fiber and integrated technologies optimized for classical
applications to address challenges of scalability in quantum technologies. 

\subsection*{Acknowledgements}

This work was funded by the National Science Foundation under Grants
PHY-1404300 and EFMA-1641094. S.C. acknowledges the F.R.S.-FNRS for
financial support. S.R. acknowledges funding by the DFG (Emmy Noether
Program). We thank Aseema Mohanty for useful comments on the manuscript. 

\subsection*{Author contributions}

S.C. conceived the idea and performed initial theoretical and experimental
groundwork. C.J., A.F. and S.R. implemented the experiment. C.J. and
A.F. performed the experiment, collected and analyzed the data. All
authors contributed to interpreting the data. C.J. and A.F. prepared
the manuscript in consultation with all authors. A.G. supervised the
project.

%\bibliographystyle{naturemag}
%\bibliography{FrequencyMultiplexing.bbl}
%\bibliography{2017-Multiplexing}

\begin{thebibliography}{100}
\expandafter\ifx\csname url\endcsname\relax
  \def\url#1{\texttt{#1}}\fi
\expandafter\ifx\csname urlprefix\endcsname\relax\def\urlprefix{URL }\fi
\providecommand{\bibinfo}[2]{#2}
\providecommand{\eprint}[2][]{\url{#2}}

\bibitem{eisaman_invited_2011-1}
\bibinfo{author}{Eisaman, M.~D.}, \bibinfo{author}{Fan, J.},
  \bibinfo{author}{Migdall, A.} \& \bibinfo{author}{Polyakov, S.~V.}
\newblock \bibinfo{title}{Invited {{Review Article}}: {{Single}}-photon sources
  and detectors}.
\newblock \emph{\bibinfo{journal}{Review of Scientific Instruments}}
  \textbf{\bibinfo{volume}{82}}, \bibinfo{pages}{071101}
  (\bibinfo{year}{2011}).

\bibitem{loredo_scalable_2016-1}
\bibinfo{author}{Loredo, J.~C.} \emph{et~al.}
\newblock \bibinfo{title}{Scalable performance in solid-state single-photon
  sources}.
\newblock \emph{\bibinfo{journal}{Optica}} \textbf{\bibinfo{volume}{3}},
  \bibinfo{pages}{433--440} (\bibinfo{year}{2016}).

\bibitem{ding_-demand_2016}
\bibinfo{author}{Ding, X.} \emph{et~al.}
\newblock \bibinfo{title}{On-demand single photons with high extraction
  efficiency and near-unity indistinguishability from a resonantly driven
  quantum dot in a micropillar}.
\newblock \emph{\bibinfo{journal}{Physical Review Letters}}
  \textbf{\bibinfo{volume}{116}}, \bibinfo{pages}{020401}
  (\bibinfo{year}{2016}).

\bibitem{somaschi_near-optimal_2016-1}
\bibinfo{author}{Somaschi, N.} \emph{et~al.}
\newblock \bibinfo{title}{Near-optimal single-photon sources in the solid
  state}.
\newblock \emph{\bibinfo{journal}{Nature Photonics}}
  \textbf{\bibinfo{volume}{10}}, \bibinfo{pages}{340--345}
  (\bibinfo{year}{2016}).

\bibitem{aharonovich_solid-state_2016}
\bibinfo{author}{Aharonovich, I.}, \bibinfo{author}{Englund, D.} \&
  \bibinfo{author}{Toth, M.}
\newblock \bibinfo{title}{Solid-state single-photon emitters}.
\newblock \emph{\bibinfo{journal}{Nature Photonics}}
  \textbf{\bibinfo{volume}{10}}, \bibinfo{pages}{631--641}
  (\bibinfo{year}{2016}).

\bibitem{ma_quantum_2012}
\bibinfo{author}{Ma, X.-S.} \emph{et~al.}
\newblock \bibinfo{title}{Quantum teleportation over 143 kilometres using
  active feed-forward}.
\newblock \emph{\bibinfo{journal}{Nature}} \textbf{\bibinfo{volume}{489}},
  \bibinfo{pages}{269--273} (\bibinfo{year}{2012}).

\bibitem{giustina_significant-loophole-free_2015}
\bibinfo{author}{Giustina, M.} \emph{et~al.}
\newblock \bibinfo{title}{Significant-{{Loophole}}-{{Free Test}} of {{Bell}}'s
  {{Theorem}} with {{Entangled Photons}}}.
\newblock \emph{\bibinfo{journal}{Physical Review Letters}}
  \textbf{\bibinfo{volume}{115}}, \bibinfo{pages}{250401}
  (\bibinfo{year}{2015}).

\bibitem{broome_photonic_2013}
\bibinfo{author}{Broome, M.~A.} \emph{et~al.}
\newblock \bibinfo{title}{Photonic boson sampling in a tunable circuit}.
\newblock \emph{\bibinfo{journal}{Science}} \textbf{\bibinfo{volume}{339}},
  \bibinfo{pages}{794--798} (\bibinfo{year}{2013}).

\bibitem{spring_boson_2013}
\bibinfo{author}{Spring, J.~B.} \emph{et~al.}
\newblock \bibinfo{title}{Boson sampling on a photonic chip}.
\newblock \emph{\bibinfo{journal}{Science}} \textbf{\bibinfo{volume}{339}},
  \bibinfo{pages}{798--801} (\bibinfo{year}{2013}).

\bibitem{shalm_strong_2015}
\bibinfo{author}{Shalm, L.~K.} \emph{et~al.}
\newblock \bibinfo{title}{Strong {{Loophole}}-{{Free Test}} of {{Local
  Realism}}}.
\newblock \emph{\bibinfo{journal}{Physical Review Letters}}
  \textbf{\bibinfo{volume}{115}}, \bibinfo{pages}{250402}
  (\bibinfo{year}{2015}).

\bibitem{kwiat_new_1995}
\bibinfo{author}{Kwiat, P.~G.} \emph{et~al.}
\newblock \bibinfo{title}{New high-intensity source of polarization-entangled
  photon pairs}.
\newblock \emph{\bibinfo{journal}{Physical Review Letters}}
  \textbf{\bibinfo{volume}{75}}, \bibinfo{pages}{4337} (\bibinfo{year}{1995}).

\bibitem{kwiat_ultrabright_1999}
\bibinfo{author}{Kwiat, P.~G.}, \bibinfo{author}{Waks, E.},
  \bibinfo{author}{White, A.~G.}, \bibinfo{author}{Appelbaum, I.} \&
  \bibinfo{author}{Eberhard, P.~H.}
\newblock \bibinfo{title}{Ultrabright source of polarization-entangled
  photons}.
\newblock \emph{\bibinfo{journal}{Physical Review A}}
  \textbf{\bibinfo{volume}{60}}, \bibinfo{pages}{R773} (\bibinfo{year}{1999}).

\bibitem{tanzilli_highly_2001}
\bibinfo{author}{Tanzilli, S.} \emph{et~al.}
\newblock \bibinfo{title}{Highly efficient photon-pair source using
  periodically poled lithium niobate waveguide}.
\newblock \emph{\bibinfo{journal}{Electronics Letters}}
  \textbf{\bibinfo{volume}{37}}, \bibinfo{pages}{26} (\bibinfo{year}{2001}).

\bibitem{fedrizzi_wavelength-tunable_2007}
\bibinfo{author}{Fedrizzi, A.}, \bibinfo{author}{Herbst, T.},
  \bibinfo{author}{Poppe, A.}, \bibinfo{author}{Jennewein, T.} \&
  \bibinfo{author}{Zeilinger, A.}
\newblock \bibinfo{title}{A wavelength-tunable fiber-coupled source of
  narrowband entangled photons}.
\newblock \emph{\bibinfo{journal}{Optics Express}}
  \textbf{\bibinfo{volume}{15}}, \bibinfo{pages}{15377--15386}
  (\bibinfo{year}{2007}).

\bibitem{rambach_sub-megahertz_2016}
\bibinfo{author}{Rambach, M.}, \bibinfo{author}{Nikolova, A.},
  \bibinfo{author}{Weinhold, T.~J.} \& \bibinfo{author}{White, A.~G.}
\newblock \bibinfo{title}{Sub-megahertz linewidth single photon source}.
\newblock \emph{\bibinfo{journal}{APL Photonics}} \textbf{\bibinfo{volume}{1}},
  \bibinfo{pages}{096101} (\bibinfo{year}{2016}).

\bibitem{nasr_ultrabroadband_2008}
\bibinfo{author}{Nasr, M.~B.} \emph{et~al.}
\newblock \bibinfo{title}{Ultrabroadband {{Biphotons Generated}} via {{Chirped
  Quasi}}-{{Phase}}-{{Matched Optical Parametric Down}}-{{Conversion}}}.
\newblock \emph{\bibinfo{journal}{Physical Review Letters}}
  \textbf{\bibinfo{volume}{100}}, \bibinfo{pages}{183601}
  (\bibinfo{year}{2008}).

\bibitem{odonnell_observation_2007}
\bibinfo{author}{O'Donnell, K.~A.} \& \bibinfo{author}{U'Ren, A.~B.}
\newblock \bibinfo{title}{Observation of ultrabroadband, beamlike parametric
  downconversion}.
\newblock \emph{\bibinfo{journal}{Optics Letters}}
  \textbf{\bibinfo{volume}{32}}, \bibinfo{pages}{817--819}
  (\bibinfo{year}{2007}).

\bibitem{reimer_integrated_2014}
\bibinfo{author}{Reimer, C.} \emph{et~al.}
\newblock \bibinfo{title}{Integrated frequency comb source of heralded single
  photons.}
\newblock \emph{\bibinfo{journal}{Optics Express}}
  \textbf{\bibinfo{volume}{22}}, \bibinfo{pages}{6535} (\bibinfo{year}{2014}).

\bibitem{reimer_generation_2016}
\bibinfo{author}{Reimer, C.} \emph{et~al.}
\newblock \bibinfo{title}{Generation of multiphoton entangled quantum states by
  means of integrated frequency combs}.
\newblock \emph{\bibinfo{journal}{Science}} \textbf{\bibinfo{volume}{351}},
  \bibinfo{pages}{1176--1180} (\bibinfo{year}{2016}).

\bibitem{ramelow_silicon-nitride_2015-1}
\bibinfo{author}{Ramelow, S.} \emph{et~al.}
\newblock \bibinfo{title}{Silicon-{{Nitride Platform}} for {{Narrowband
  Entangled Photon Generation}}}.
\newblock \emph{\bibinfo{journal}{arXiv:1508.04358 [physics,
  physics:quant-ph]}}  (\bibinfo{year}{2015}).
\newblock \eprint{1508.04358}.

\bibitem{migdall_tailoring_2002}
\bibinfo{author}{Migdall, A.~L.}, \bibinfo{author}{Branning, D.} \&
  \bibinfo{author}{Castelletto, S.}
\newblock \bibinfo{title}{Tailoring single-photon and multiphoton probabilities
  of a single-photon on-demand source}.
\newblock \emph{\bibinfo{journal}{Physical Review A}}
  \textbf{\bibinfo{volume}{66}}, \bibinfo{pages}{053805}
  (\bibinfo{year}{2002}).

\bibitem{jeffrey_towards_2004}
\bibinfo{author}{Jeffrey, E.}, \bibinfo{author}{Peters, N.~A.} \&
  \bibinfo{author}{Kwiat, P.~G.}
\newblock \bibinfo{title}{Towards a periodic deterministic source of arbitrary
  single-photon states}.
\newblock \emph{\bibinfo{journal}{New Journal of Physics}}
  \textbf{\bibinfo{volume}{6}}, \bibinfo{pages}{100--100}
  (\bibinfo{year}{2004}).

\bibitem{pittman_single_2002}
\bibinfo{author}{Pittman, T.~B.}, \bibinfo{author}{Jacobs, B.~C.} \&
  \bibinfo{author}{Franson, J.~D.}
\newblock \bibinfo{title}{Single photons on pseudodemand from stored parametric
  down-conversion}.
\newblock \emph{\bibinfo{journal}{Physical Review A}}
  \textbf{\bibinfo{volume}{66}}, \bibinfo{pages}{042303}
  (\bibinfo{year}{2002}).

\bibitem{shapiro_-demand_2007}
\bibinfo{author}{Shapiro, J.~H.} \& \bibinfo{author}{Wong, F.~N.}
\newblock \bibinfo{title}{On-demand single-photon generation using a modular
  array of parametric downconverters with electro-optic polarization controls}.
\newblock \emph{\bibinfo{journal}{Optics Letters}}
  \textbf{\bibinfo{volume}{32}}, \bibinfo{pages}{2698--2700}
  (\bibinfo{year}{2007}).

\bibitem{mower_efficient_2011}
\bibinfo{author}{Mower, J.} \& \bibinfo{author}{Englund, D.}
\newblock \bibinfo{title}{Efficient generation of single and entangled photons
  on a silicon photonic integrated chip}.
\newblock \emph{\bibinfo{journal}{Physical Review A}}
  \textbf{\bibinfo{volume}{84}}, \bibinfo{pages}{052326}
  (\bibinfo{year}{2011}).

\bibitem{christ_limits_2012-2}
\bibinfo{author}{Christ, A.} \& \bibinfo{author}{Silberhorn, C.}
\newblock \bibinfo{title}{Limits on the deterministic creation of pure
  single-photon states using parametric down-conversion}.
\newblock \emph{\bibinfo{journal}{Physical Review A}}
  \textbf{\bibinfo{volume}{85}}, \bibinfo{pages}{023829}
  (\bibinfo{year}{2012}).

\bibitem{ma_experimental_2011}
\bibinfo{author}{Ma, X.-s.}, \bibinfo{author}{Zotter, S.},
  \bibinfo{author}{Kofler, J.}, \bibinfo{author}{Jennewein, T.} \&
  \bibinfo{author}{Zeilinger, A.}
\newblock \bibinfo{title}{Experimental generation of single photons via active
  multiplexing}.
\newblock \emph{\bibinfo{journal}{Physical Review A}}
  \textbf{\bibinfo{volume}{83}}, \bibinfo{pages}{043814}
  (\bibinfo{year}{2011}).

\bibitem{broome_reducing_2011}
\bibinfo{author}{Broome, M.~A.}, \bibinfo{author}{Almeida, M.~P.},
  \bibinfo{author}{Fedrizzi, A.} \& \bibinfo{author}{White, A.~G.}
\newblock \bibinfo{title}{Reducing multi-photon rates in pulsed down-conversion
  by temporal multiplexing}.
\newblock \emph{\bibinfo{journal}{Optics Express}}
  \textbf{\bibinfo{volume}{19}}, \bibinfo{pages}{22698} (\bibinfo{year}{2011}).

\bibitem{collins_integrated_2013}
\bibinfo{author}{Collins, M.~J.} \emph{et~al.}
\newblock \bibinfo{title}{Integrated spatial multiplexing of heralded
  single-photon sources.}
\newblock \emph{\bibinfo{journal}{Nature communications}}
  \textbf{\bibinfo{volume}{4}}, \bibinfo{pages}{2582} (\bibinfo{year}{2013}).

\bibitem{mendoza_active_2016}
\bibinfo{author}{Mendoza, G.~J.} \emph{et~al.}
\newblock \bibinfo{title}{Active temporal and spatial multiplexing of photons}.
\newblock \emph{\bibinfo{journal}{Optica}} \textbf{\bibinfo{volume}{3}},
  \bibinfo{pages}{127} (\bibinfo{year}{2016}).

\bibitem{xiong_active_2016}
\bibinfo{author}{Xiong, C.} \emph{et~al.}
\newblock \bibinfo{title}{Active temporal multiplexing of indistinguishable
  heralded single photons}.
\newblock \emph{\bibinfo{journal}{Nature Communications}}
  \textbf{\bibinfo{volume}{7}}, \bibinfo{pages}{10853} (\bibinfo{year}{2016}).

\bibitem{kaneda_time-multiplexed_2015-1}
\bibinfo{author}{Kaneda, F.} \emph{et~al.}
\newblock \bibinfo{title}{Time-multiplexed heralded single-photon source}.
\newblock \emph{\bibinfo{journal}{Optica}} \textbf{\bibinfo{volume}{2}},
  \bibinfo{pages}{1010--1013} (\bibinfo{year}{2015}).

\bibitem{francis-jones_all-fiber_2016}
\bibinfo{author}{Francis-Jones, R. J.~A.}, \bibinfo{author}{Hoggarth, R.~A.} \&
  \bibinfo{author}{Mosley, P.~J.}
\newblock \bibinfo{title}{All-fiber multiplexed source of high-purity single
  photons}.
\newblock \emph{\bibinfo{journal}{Optica}} \textbf{\bibinfo{volume}{3}},
  \bibinfo{pages}{1270--1273} (\bibinfo{year}{2016}).

\bibitem{inoue_tunable_1994}
\bibinfo{author}{Inoue, K.}
\newblock \bibinfo{title}{Tunable and selective wavelength conversion using
  fiber four-wave mixing with two pump lights}.
\newblock \emph{\bibinfo{journal}{IEEE Photonics Technology Letters}}
  \textbf{\bibinfo{volume}{6}}, \bibinfo{pages}{1451--1453}
  (\bibinfo{year}{1994}).

\bibitem{mckinstrie_translation_2005}
\bibinfo{author}{McKinstrie, C.~J.}, \bibinfo{author}{Harvey, J.~D.},
  \bibinfo{author}{Radic, S.} \& \bibinfo{author}{Raymer, M.~G.}
\newblock \bibinfo{title}{Translation of quantum states by four-wave mixing in
  fibers}.
\newblock \emph{\bibinfo{journal}{Optics Express}}
  \textbf{\bibinfo{volume}{13}}, \bibinfo{pages}{9131} (\bibinfo{year}{2005}).

\bibitem{mcguinness_quantum_2010}
\bibinfo{author}{McGuinness, H.~J.}, \bibinfo{author}{Raymer, M.~G.},
  \bibinfo{author}{McKinstrie, C.~J.} \& \bibinfo{author}{Radic, S.}
\newblock \bibinfo{title}{Quantum {{Frequency Translation}} of
  {{Single}}-{{Photon States}} in a {{Photonic Crystal Fiber}}}.
\newblock \emph{\bibinfo{journal}{Physical Review Letters}}
  \textbf{\bibinfo{volume}{105}}, \bibinfo{pages}{093604}
  (\bibinfo{year}{2010}).

\bibitem{li_efficient_2016}
\bibinfo{author}{Li, Q.}, \bibinfo{author}{Davan{\c c}o, M.} \&
  \bibinfo{author}{Srinivasan, K.}
\newblock \bibinfo{title}{Efficient and low-noise single-photon-level frequency
  conversion interfaces using silicon nanophotonics}.
\newblock \emph{\bibinfo{journal}{Nature Photonics}}
  \textbf{\bibinfo{volume}{10}}, \bibinfo{pages}{406--414}
  (\bibinfo{year}{2016}).

\bibitem{farsi_low-noise_2015}
\bibinfo{author}{Farsi, A.}, \bibinfo{author}{Clemmen, S.},
  \bibinfo{author}{Ramelow, S.} \& \bibinfo{author}{Gaeta, A.~L.}
\newblock \bibinfo{title}{Low-noise quantum frequency translation of single
  photons}.
\newblock \bibinfo{pages}{FM3A.4} (\bibinfo{publisher}{{OSA}},
  \bibinfo{year}{2015}).

\bibitem{clemmen_ramsey_2016}
\bibinfo{author}{Clemmen, S.}, \bibinfo{author}{Farsi, A.},
  \bibinfo{author}{Ramelow, S.} \& \bibinfo{author}{Gaeta, A.~L.}
\newblock \bibinfo{title}{Ramsey interference with single photons}.
\newblock \emph{\bibinfo{journal}{Physical Review Letters}}
  \textbf{\bibinfo{volume}{117}}, \bibinfo{pages}{223601}
  (\bibinfo{year}{2016}).

\bibitem{joshi_frequency_2016}
\bibinfo{author}{Joshi, C.}, \bibinfo{author}{Farsi, A.},
  \bibinfo{author}{Ramelow, S.}, \bibinfo{author}{Clemmen, S.} \&
  \bibinfo{author}{Gaeta, A.~L.}
\newblock \bibinfo{title}{Frequency multiplexing for quasi-deterministic
  heralded single-photon sources}.
\newblock In \emph{\bibinfo{booktitle}{Conference on {{Lasers}} and
  {{Electro}}-{{Optics}} (2016), paper {{FTh1C}}.2}}, \bibinfo{pages}{FTh1C.2}
  (\bibinfo{publisher}{{Optical Society of America}}, \bibinfo{year}{2016}).

\bibitem{roslund_wavelength-multiplexed_2014}
\bibinfo{author}{Roslund, J.}, \bibinfo{author}{{de Ara{\'u}jo}, R.~M.},
  \bibinfo{author}{Jiang, S.}, \bibinfo{author}{Fabre, C.} \&
  \bibinfo{author}{Treps, N.}
\newblock \bibinfo{title}{Wavelength-multiplexed quantum networks with
  ultrafast frequency combs}.
\newblock \emph{\bibinfo{journal}{Nature Photonics}}
  \textbf{\bibinfo{volume}{8}}, \bibinfo{pages}{109--112}
  (\bibinfo{year}{2014}).

\bibitem{humphreys_continuous-variable_2014}
\bibinfo{author}{Humphreys, P.~C.} \emph{et~al.}
\newblock \bibinfo{title}{Continuous-variable quantum computing in optical
  time-frequency modes using quantum memories}.
\newblock \emph{\bibinfo{journal}{Physical Review Letters}}
  \textbf{\bibinfo{volume}{113}}, \bibinfo{pages}{130502}
  (\bibinfo{year}{2014}).

\bibitem{sinclair_spectral_2014}
\bibinfo{author}{Sinclair, N.} \emph{et~al.}
\newblock \bibinfo{title}{Spectral {{Multiplexing}} for {{Scalable Quantum
  Photonics}} using an {{Atomic Frequency Comb Quantum Memory}} and
  {{Feed}}-{{Forward Control}}}.
\newblock \emph{\bibinfo{journal}{Physical Review Letters}}
  \textbf{\bibinfo{volume}{113}}, \bibinfo{pages}{053603}
  (\bibinfo{year}{2014}).

\bibitem{lukens_frequency-encoded_2017}
\bibinfo{author}{Lukens, J.~M.} \& \bibinfo{author}{Lougovski, P.}
\newblock \bibinfo{title}{Frequency-encoded photonic qubits for scalable
  quantum information processing}.
\newblock \emph{\bibinfo{journal}{Optica}} \textbf{\bibinfo{volume}{4}},
  \bibinfo{pages}{8--16} (\bibinfo{year}{2017}).

\bibitem{puigibert_heralded_2017}
\bibinfo{author}{Puigibert, M.~G.} \emph{et~al.}
\newblock \bibinfo{title}{Heralded single photons based on spectral
  multiplexing and feed-forward control}.
\newblock \emph{\bibinfo{journal}{arXiv:1703.02068 [quant-ph]}}
  (\bibinfo{year}{2017}).
\newblock \eprint{1703.02068}.

\bibitem{bonneau_effect_2015}
\bibinfo{author}{Bonneau, D.}, \bibinfo{author}{Mendoza, G.~J.},
  \bibinfo{author}{O'Brien, J.~L.} \& \bibinfo{author}{Thompson, M.~G.}
\newblock \bibinfo{title}{Effect of loss on multiplexed single-photon sources}.
\newblock \emph{\bibinfo{journal}{New Journal of Physics}}
  \textbf{\bibinfo{volume}{17}}, \bibinfo{pages}{043057}
  (\bibinfo{year}{2015}).

\bibitem{brown_correlation_1956}
\bibinfo{author}{Brown, R.~H.} \& \bibinfo{author}{Twiss, R.~Q.}
\newblock \bibinfo{title}{Correlation between {{Photons}} in two {{Coherent
  Beams}} of {{Light}}}.
\newblock \emph{\bibinfo{journal}{Nature}} \textbf{\bibinfo{volume}{177}},
  \bibinfo{pages}{27--29} (\bibinfo{year}{1956}).

\bibitem{ramelow_highly_2013}
\bibinfo{author}{Ramelow, S.} \emph{et~al.}
\newblock \bibinfo{title}{Highly efficient heralding of entangled single
  photons}.
\newblock \emph{\bibinfo{journal}{Optics Express}}
  \textbf{\bibinfo{volume}{21}}, \bibinfo{pages}{6707} (\bibinfo{year}{2013}).

\bibitem{helt_spontaneous_2010}
\bibinfo{author}{Helt, L.~G.}, \bibinfo{author}{Yang, Z.},
  \bibinfo{author}{Liscidini, M.} \& \bibinfo{author}{Sipe, J.~E.}
\newblock \bibinfo{title}{Spontaneous four-wave mixing in microring
  resonators}.
\newblock \emph{\bibinfo{journal}{Optics Letters}}
  \textbf{\bibinfo{volume}{35}}, \bibinfo{pages}{3006--3008}
  (\bibinfo{year}{2010}).

\bibitem{azuma_all-photonic_2015}
\bibinfo{author}{Azuma, K.}, \bibinfo{author}{Tamaki, K.} \&
  \bibinfo{author}{Lo, H.-K.}
\newblock \bibinfo{title}{All-photonic quantum repeaters}.
\newblock \emph{\bibinfo{journal}{Nature Communications}}
  \textbf{\bibinfo{volume}{6}}, \bibinfo{pages}{ncomms7787}
  (\bibinfo{year}{2015}).

\bibitem{agha_low-noise_2012}
\bibinfo{author}{Agha, I.}, \bibinfo{author}{Davan{\c c}o, M.},
  \bibinfo{author}{Thurston, B.} \& \bibinfo{author}{Srinivasan, K.}
\newblock \bibinfo{title}{Low-noise chip-based frequency conversion by
  four-wave-mixing {{Bragg}} scattering in {{SiN}}{\textsubscript{x}}
  waveguides}.
\newblock \emph{\bibinfo{journal}{Optics Letters}}
  \textbf{\bibinfo{volume}{37}}, \bibinfo{pages}{2997--2999}
  (\bibinfo{year}{2012}).

\end{thebibliography}

\newpage

\clearpage
\setcounter{section}{0}
\renewcommand\thefigure{S\arabic{figure}}     
\setcounter{figure}{0}
\renewcommand\thetable{S\arabic{table}}
\setcounter{table}{0}

\part*{Supplementary Information}

\section{Bragg scattering four-wave mixing}

\pagenumbering{arabic} 

\begin{figure}[t]
\vspace{-60pt}

\subfloat[]{\includegraphics[width=0.95\textwidth]{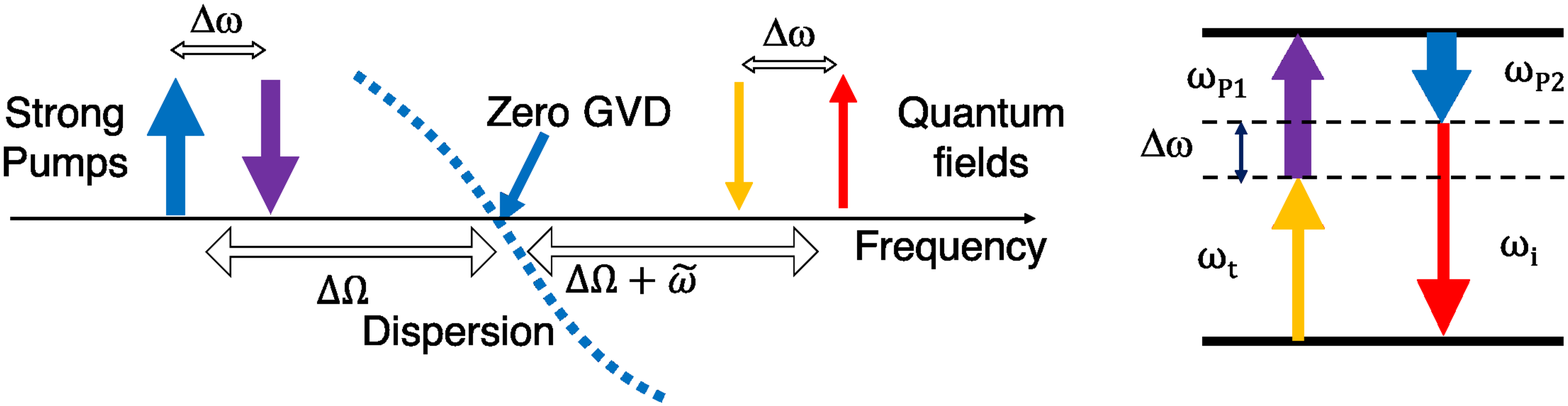}

}

\subfloat[]{

\includegraphics[width=0.45\textwidth]{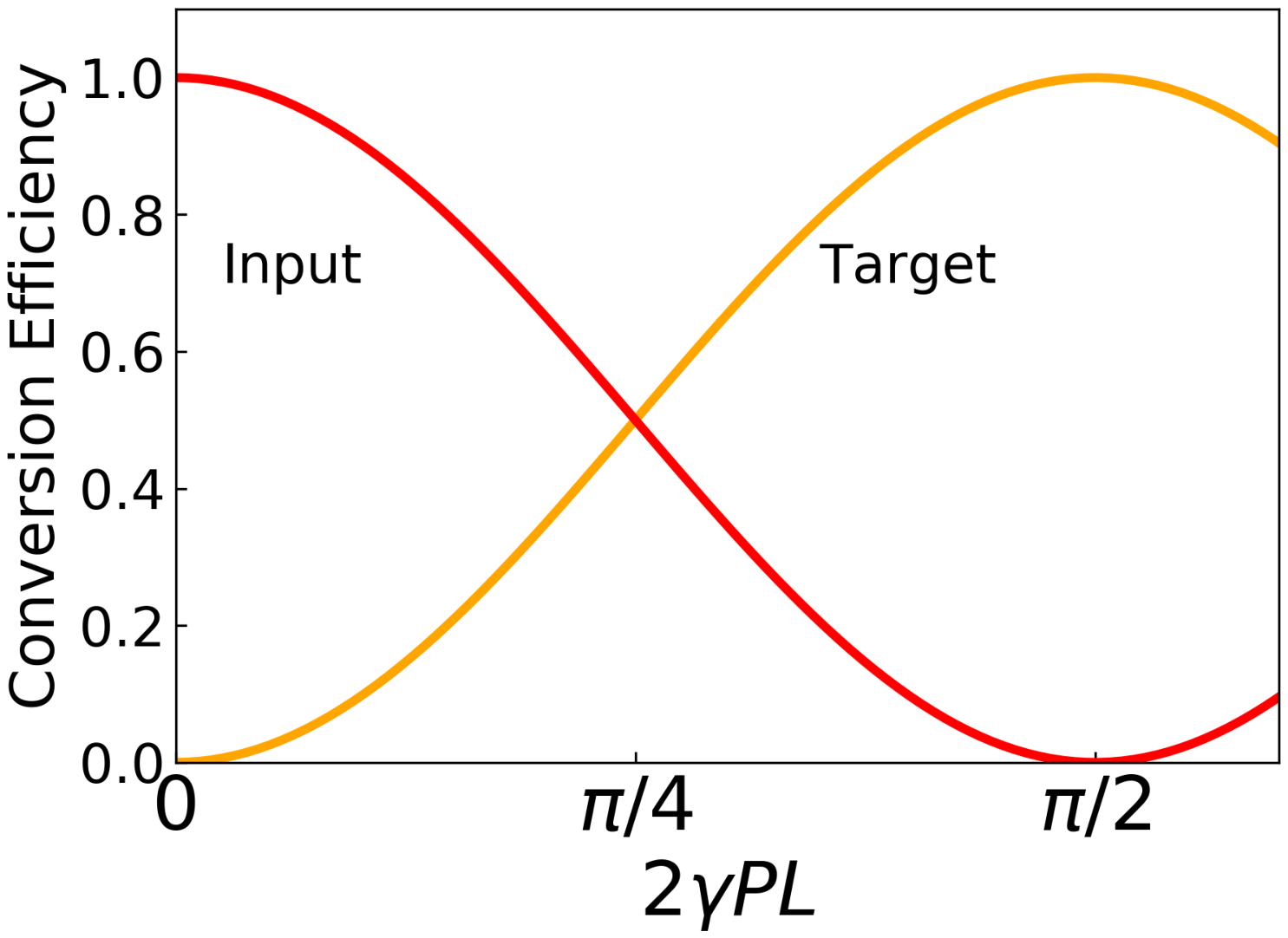}}\subfloat[]{

\includegraphics[width=0.5\textwidth]{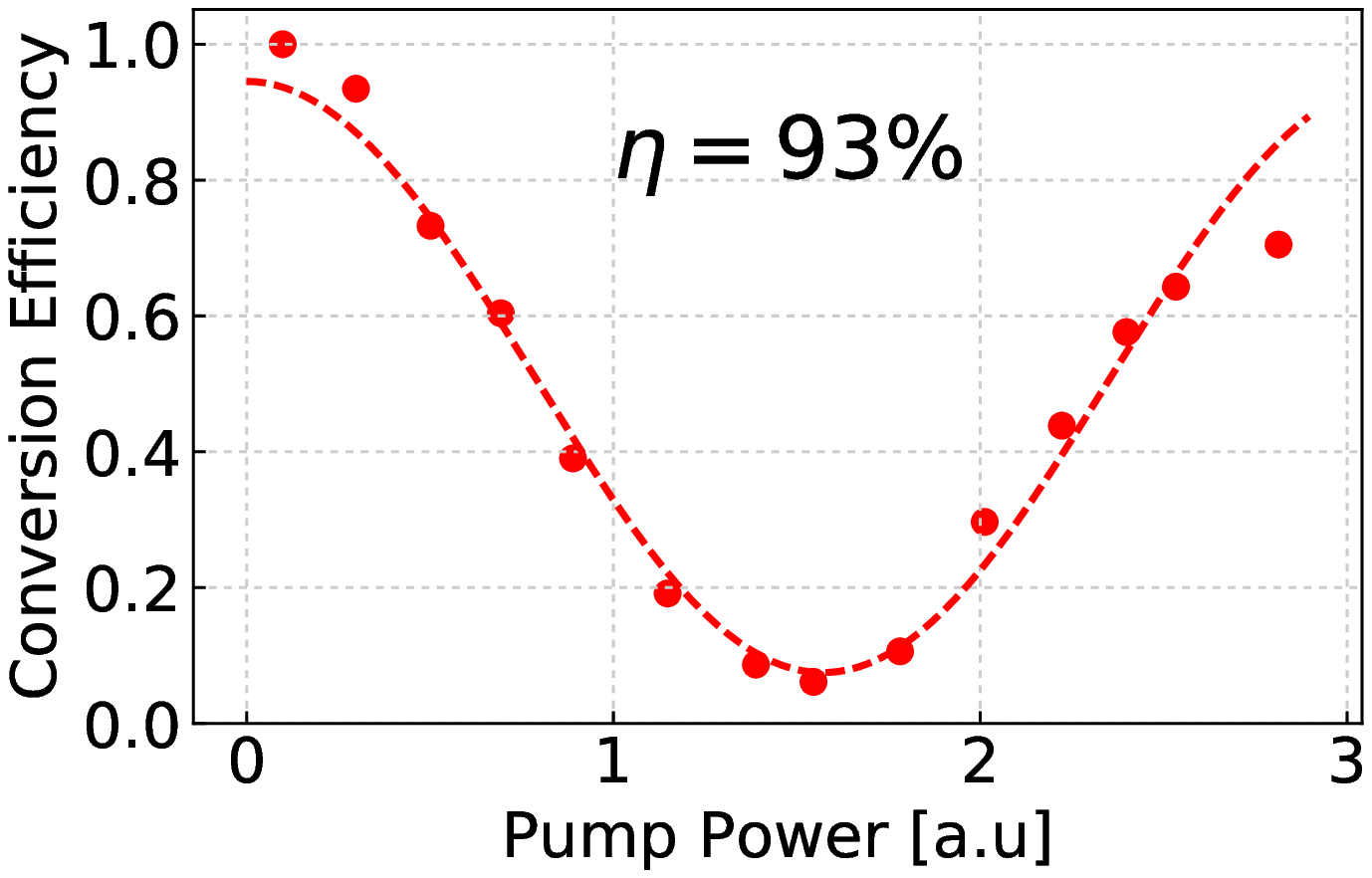}}\vfill{}
\subfloat[]{\includegraphics[width=0.5\textwidth]{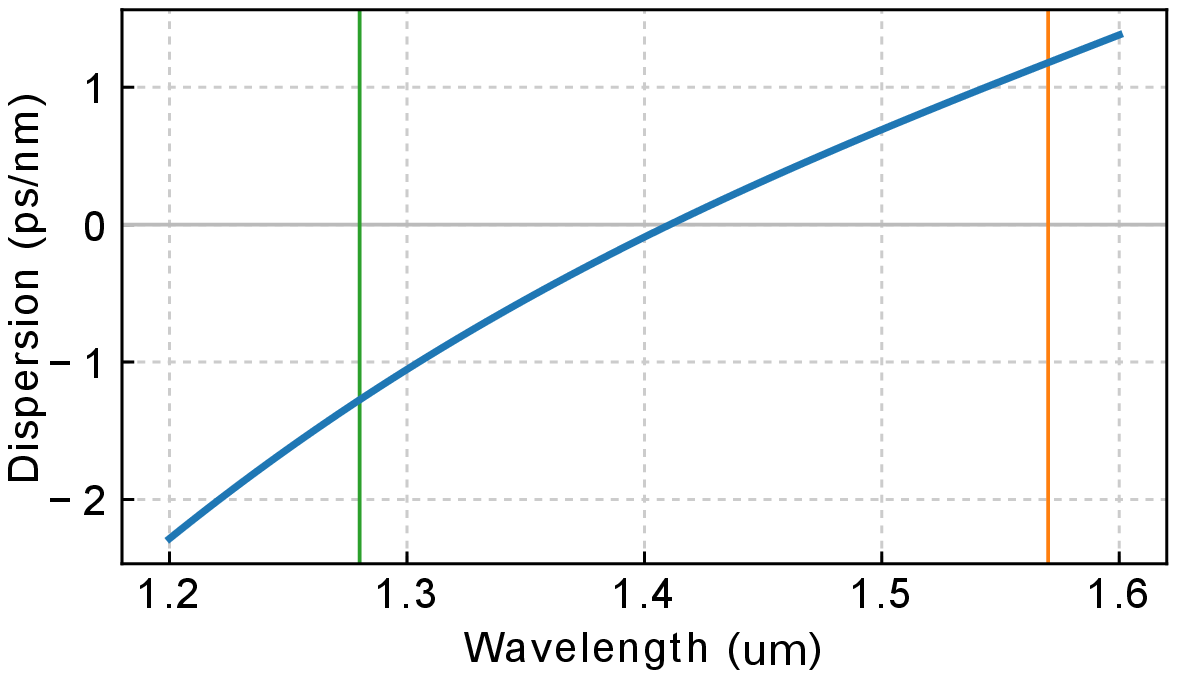}}\subfloat[]{\includegraphics[width=0.5\textwidth]{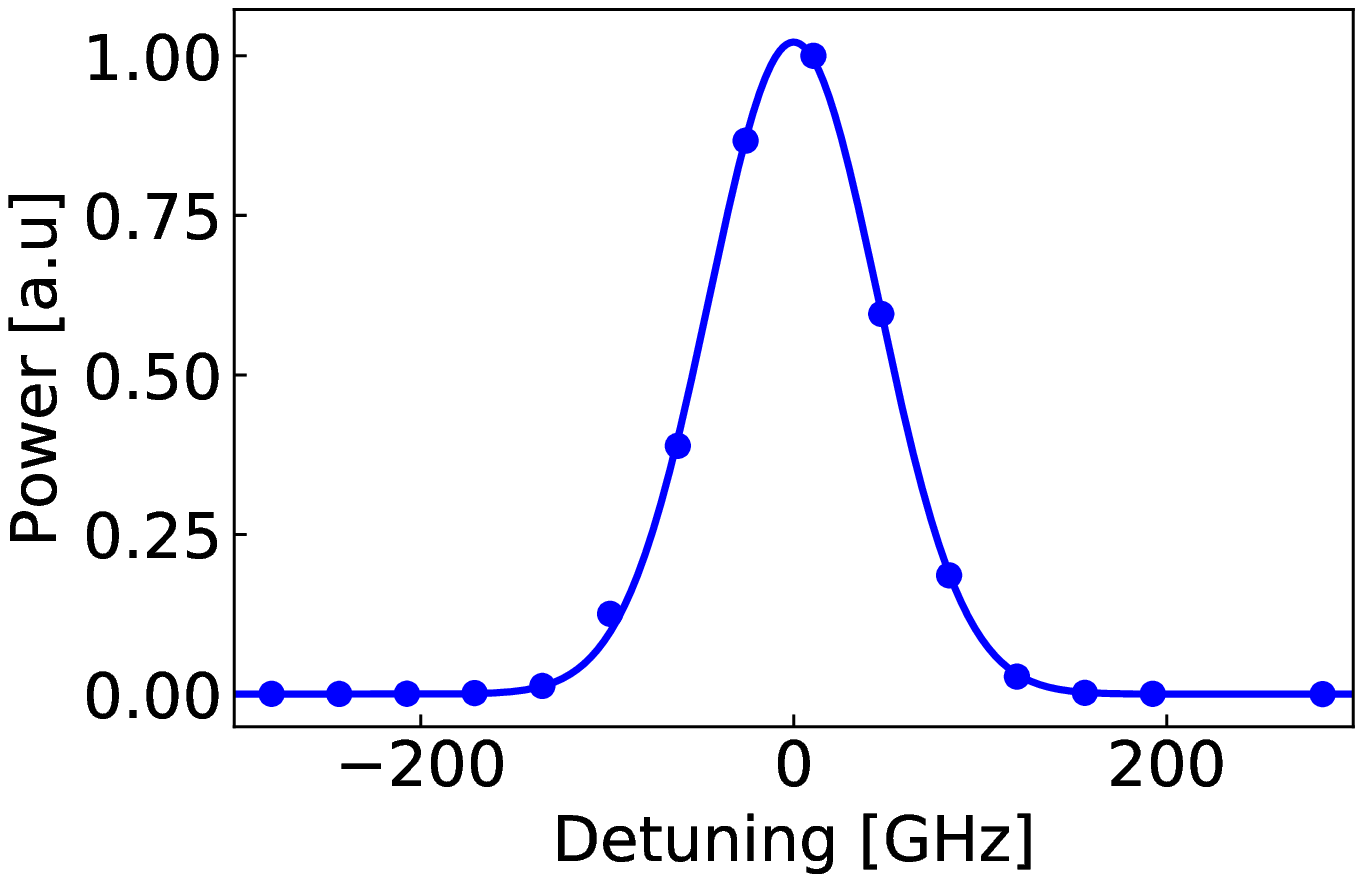}

}

\caption{\label{fig:braggconversion} a) BS-FWM: Two strong pumps $\omega_{P1}$
and $\omega_{P2}$ drive the interaction between two fields $\omega_{s}$
and $\omega_{i}$, where the separation $\Delta\omega=\omega_{P1}-\omega_{P2}$
determines the frequency shift between the signal and idler fields.
Phase matching is achieved by symmetric placement of the pumps, and
the signal, idler fields about the ZDW of the interaction medium.
b) Bragg scattering conversion efficiency as function of interaction
strength $2\gamma P$, assuming perfect phase matching. Complete conversion
is achieved when the interaction strength equals $\pi/2$. c) Measured
conversion efficiency via depletion of the signal. The conversion
efficiency is measured to be $93\%$ , limited by fluctuations in
the pump power. d) Dispersion profile of the dispersion shifted fiber
used in our implementation of BS-FWM. e) Measured bandwidth of the
filtering setup centered at the target frequency. }

\end{figure}

Four wave Mixing Bragg scattering is a parametric process driven by
two strong pump fields at $\omega_{P1}$ and $\omega_{P2}$ separated
by $\Delta\omega=\omega_{P1}-\omega_{P2}$ that can convert an input
field $\omega_{i}$ to $\omega_{t}=\omega_{i}-\Delta\omega$ (see
Figure \ref{fig:braggconversion}a). The efficiency of the conversion
is $\eta=\frac{k^{2}}{k^{2}+\kappa^{2}}\sin(\sqrt{k^{2}+\kappa^{2}}L)$
where $\kappa=2P\gamma$ is the nonlinear strength, $k=\frac{1}{2}(\beta_{P1}-\beta_{P2}+\beta_{t}-\beta_{i})$
is the phase mismatch, $L$ the interaction length, $P$ is the power
in each pump, $\gamma$ is the nonlinear coefficient of the medium,
and $\beta_{P1,P2,t,i}$ the propagation vector for each participating
fields. At the end of the interaction, the signal intensity is depleted
by a factor of $1-\eta$. For a perfectly phase-matched process ($k=0$),
complete conversion is achieved when the interaction strength $2\gamma PL$
equals $\pi/2$ (figure \ref{fig:braggconversion}b). Tunable frequency
conversion can be achieved by tuning the separation $\Delta\omega$
between the two strong pumps. Phase matching is achieved by symmetric
placement of the pumps and the input and target fields around the
zero-dispersion wavelength ($\beta^{(2)}=0$ ). The acceptance bandwidth
$\Delta\nu_{BS}$ of this process is determined by higher order dispersion. 

The setup used for this experiment is similar to \cite{farsi_low-noise_2015,clemmen_ramsey_2016}.
We use a 100-m long dispersion shifted fiber with the zero dispersion
wavelength $\lambda_{zgvd}=1405$ nm as the nonlinear medium. The
fiber is cooled using liquid nitrogen in order to remove spontaneous
Raman noise. Pumps are generated by temperature stabilized distributed
feed-back laser diodes and combined together using DWDMs. Within the
bandwidth of the filter on the target channel, we measure a background
of $3\times10^{-3}$ photons per pump pulse duration. We achieve 93\%
conversion efficiency with this setup, as shown in \ref{fig:braggconversion}c.
The conversion efficiency is limited by the fluctuations in the pump
power due to random triggering set by the CW nature of the photon
source. With a fixed periodic trigger rate, we measure a conversion
rate of $97\%$ . Our filtering bandwidth at the target output is
set to 100 GHz (figure \ref{fig:braggconversion}e), which is less
than the acceptance bandwidth of the process ($\Delta\nu_{BS}=160$
GHz). 

\begin{table*}[!t]
\hspace{-60pt}

\begin{tabular}{|>{\centering}p{65pt}|>{\centering}p{60pt}|>{\centering}p{50pt}|>{\centering}p{40pt}|>{\centering}p{65pt}|>{\centering}p{65pt}|c|>{\centering}p{50pt}|>{\centering}p{50pt}|}
\hline 
Reference & Multiplexing scheme & Platform & \#Modes N & Single-photon detection rate (kHz) & Enhancement factor & $g^{(2)}(0)$ & Heralding efficiency & Max. switching speed\tabularnewline
\hline 
\hline 
This work & Frequency & Fiber  & 3 & 23 $(46^{\dagger})$ & 2.2 & 0.07 & 2.3\%

($4.6\%^{\dagger}$) & 5 MHz\tabularnewline
\hline 
Kaneda et al. \cite{kaneda_time-multiplexed_2015-1} & \multicolumn{1}{c|}{Time} & Free space & \multicolumn{1}{c|}{30} & 11 ($19.3^{\dagger}$) & 6 & 0.48 & 22\% 

(38.6\%$^{\dagger}$) & 50 kHz\tabularnewline
\hline 
Xiong et al. \cite{xiong_active_2016} & Time & Fiber  & 4 & 0.6 & 2 & ... & ... & ...\tabularnewline
\hline 
Mendoza et al. \cite{mendoza_active_2016} & Space-Time & Fiber  & 8 & 0.4 & 1.8 & ... & $\leq0.5\%^{*}$ & 500 kHz\tabularnewline
\hline 
Puigibert et al.\cite{puigibert_heralded_2017} & Frequency & Fiber & 3 & 0.4 & 1 & 0.06 & ... & ...\tabularnewline
\hline 
Francis-Jones et al.\cite{francis-jones_all-fiber_2016} & Space & Fiber & 2 & 0.6 & 1.4 & 0.05 & $...$ & ...\tabularnewline
\hline 
Ma et al.\cite{ma_experimental_2011} & Space & Free space & 4 & 0.7 & 1.4 & 0.1 & $...$ & 15 MHz\tabularnewline
\hline 
Collins et al. \cite{collins_integrated_2013} & Space & Integrated & 2 & 0.02 & 1.6 & 0.2 & 1\% & 1 MHz\tabularnewline
\hline 
\end{tabular}
\begin{raggedright}
$^{\dagger}$generation rates are estimated after correcting for detection
loss
\par\end{raggedright}
\begin{raggedright}
$^{*}$estimated from reported data 
\par\end{raggedright}
\caption{\label{tab:Comparison-of-the-1}Comparison of the performance of the
frequency multiplexed source with other demonstrations of multiplexed
single photon sources using various schemes and platforms. }
\end{table*}

\newpage

\section{Theory of Multiplexed sources}

We adapt the analysis presented in References \cite{christ_limits_2012-2,bonneau_effect_2015}.
We assume that the SPDC source generates a two-mode squeezed state: 

\begin{align}
|\psi\rangle &= \sqrt{1 - |\xi|^2}\sum_{n = 0}^{n =\infty}\xi^{n}|n_s,n_i\rangle
\end{align}
where  $\xi$ is the squeezing parameter.

\begin{figure*}[p]
\includegraphics[width=0.5\paperwidth]{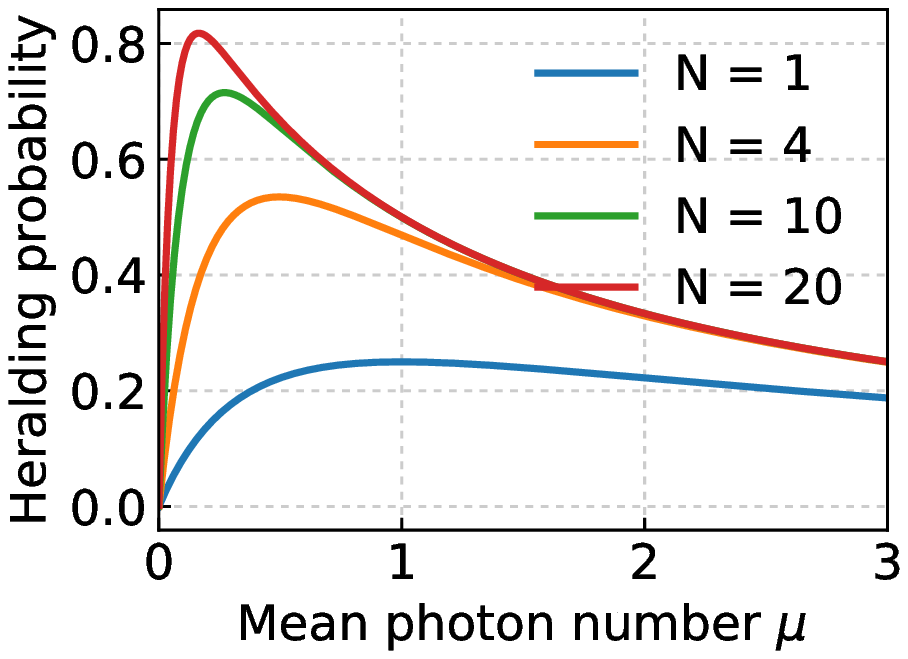}\caption{\label{fig:Heralding-probability-for}Heralding probability for $N$
multiplexed sources as a function of the mean photon number $\mu$
in the signal and idler modes, for a lossless switching network. Optimal
$\mu$ reduces as $N$ increases. }
\end{figure*}

Assuming the use of bucket (non-photon number resolving detectors), the total heralding probability is obtained by summing over the probability of an n-photon fock state registering a 'click' on the detector, given by $[1-(1-\eta_{h})^n]$, over all n. 
\begin{align}
p_h &= \frac{\eta_{h}|\xi|^2}{1 - (1-\eta_{h})|\xi|^2}
\end{align}
where $\eta_h$ is the net effciency on the heralding arm. \\
The conditional probability of detecting a single photon on the heralded arm is then given by: 
\begin{align}
p_s &= \frac{1}{p_h}\eta_d\eta_h|\xi|^2(1 - |\xi|^2)
\end{align}
where $\eta_d$ is the net efficiency on the heralded photon arm.
For $N$ multiplexed sources, the probability that a heralding photon is registered in at least one of the $N$ sources is given as: 
\begin{align}
p^{mux}_h(N) = 1 - (1-p_h)^N
\end{align}
If the net efficiency of the switching network is $\eta_{switch}$, 
the corresponding total probability that a single photon is heralded at the ouput is given by combining Eqns. 4 and 5: 
\begin{align}
p^{mux}_s(N) = \eta_{switch}\times p_{s} \times p^{mux}_{h}(N)
\end{align}
Here the loss factor $\eta_{switch}$ for the fixed loss and log tree architectures is given as: 
\begin{align}
\eta^{fixed-loss}_{switch} &= \eta_s  \\
\eta^{log-tree}_{switch} &= \eta_s^{\lceil \log_2 N \rceil} 
\end{align}
where $\eta_s$ is the switching effciency per switch. 
The expressions for the multi-pass scheme are slightly more involved. We assume optimization similar to that in Ref. \cite{kaneda_time-multiplexed_2015-1} where losses are minimized by routing photons from the the last heralded slot in case of temporal multiplexing. The total heralding probability in this case is given by: 
\begin{align}
p^{mux}_{s} = \sum_{j = 1}^{N}(1-p_h)^{N-j}p_{h}\times p_s \eta^{N-j}_s
\end{align}
In Eqn [8] the first term is the probability that no photon was heralded in the last $N-j$ slots,and the corresponding switching losses are equal to $\eta_s^{N-j}$. We note that we do not consider the cases where multi-photon emission is detected as a single photon event 
due to switching losses. 
In order to isolate effects of switching losses, we assume all other components including detection to be perfect $(\eta_d = 1, \eta_h = 1)$. \\
While determining the scaling performance of the multiplexed sources for various $N$, we optimize the mean photon number (or equivalently the squeezing parameter) for each $N$. 
The squeezing parameter $\xi$ is related to the mean photon number in the signal and idler modes as: 
\begin{align}
\mu = \frac{|\xi|^2}{1-|\xi|^2}
\end{align}
For small heralding probability $p_h$, the probability that atleast one multiplexed source triggers is $~Np_h$. As $N$ increases, optimal performance is achieved for lower mean photon number $\mu$, as shown in Fig. \ref{fig:Heralding-probability-for}. Therefore, to obtain the scaling performance of the schemes we optimize  $\mu$ for each $N$, maintaining this $\mu$ across schemes. \\
Finally, the conditional multi-photon probability for a given squeezing parameter is (ignoring switching losses): 
\begin{align}
\begin{split}
p_{multi} &= |\xi|^4 \\
&= \left(\frac{\mu}{1 + \mu}\right)^2
\end{split}
\end{align}
As the mean photon number is reduced for increasing $N$, the multi-photon noise correspondingly reduces for large $N$.

\section{Source Characterization}

\begin{figure*}
\includegraphics[width=0.5\paperwidth]{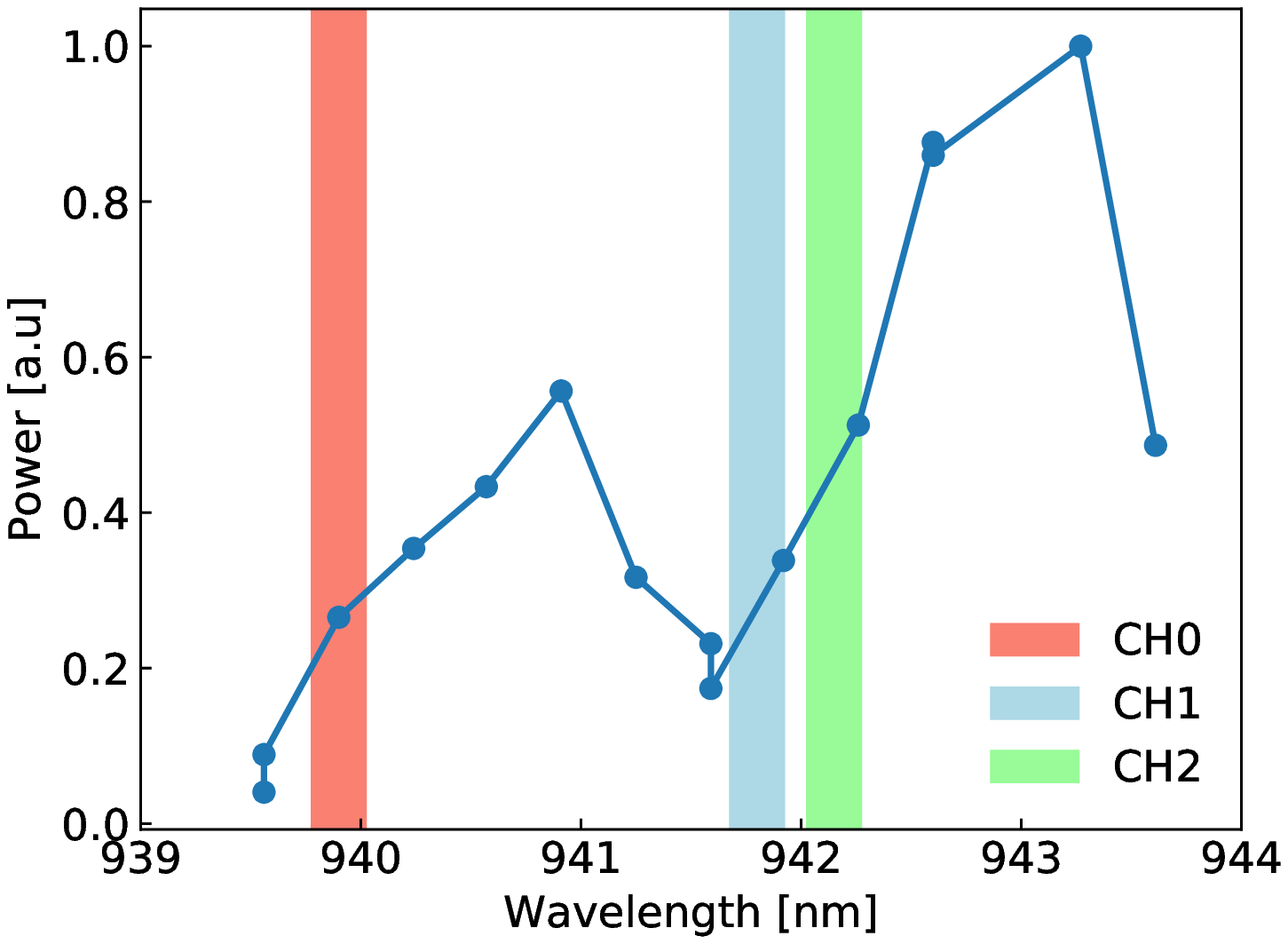}\caption{\label{fig:Spectrum-of-the}Spectrum of the heralding photons measured
using a single photon spectrometer. The heralding photons are filtered
into 100GHz wide channels using reflecting Bragg gratings. The highlighted
regions reflect the corresponding heralded channels: CH0, CH1, CH2. }
\end{figure*}

We characterize the spectrum of our idler (heralding) photons using
a single photon spectrometer (Ocean Optics). The results are shown
in Fig. \ref{fig:Spectrum-of-the}. The heralding photons are filtered
into 100 GHz wide channels using reflecting Bragg gratings. The highlighted
regions reflect the corresponding heralded channels. 

\section{characterization of system efficiency}

\begin{figure*}
\includegraphics[width=0.5\paperwidth]{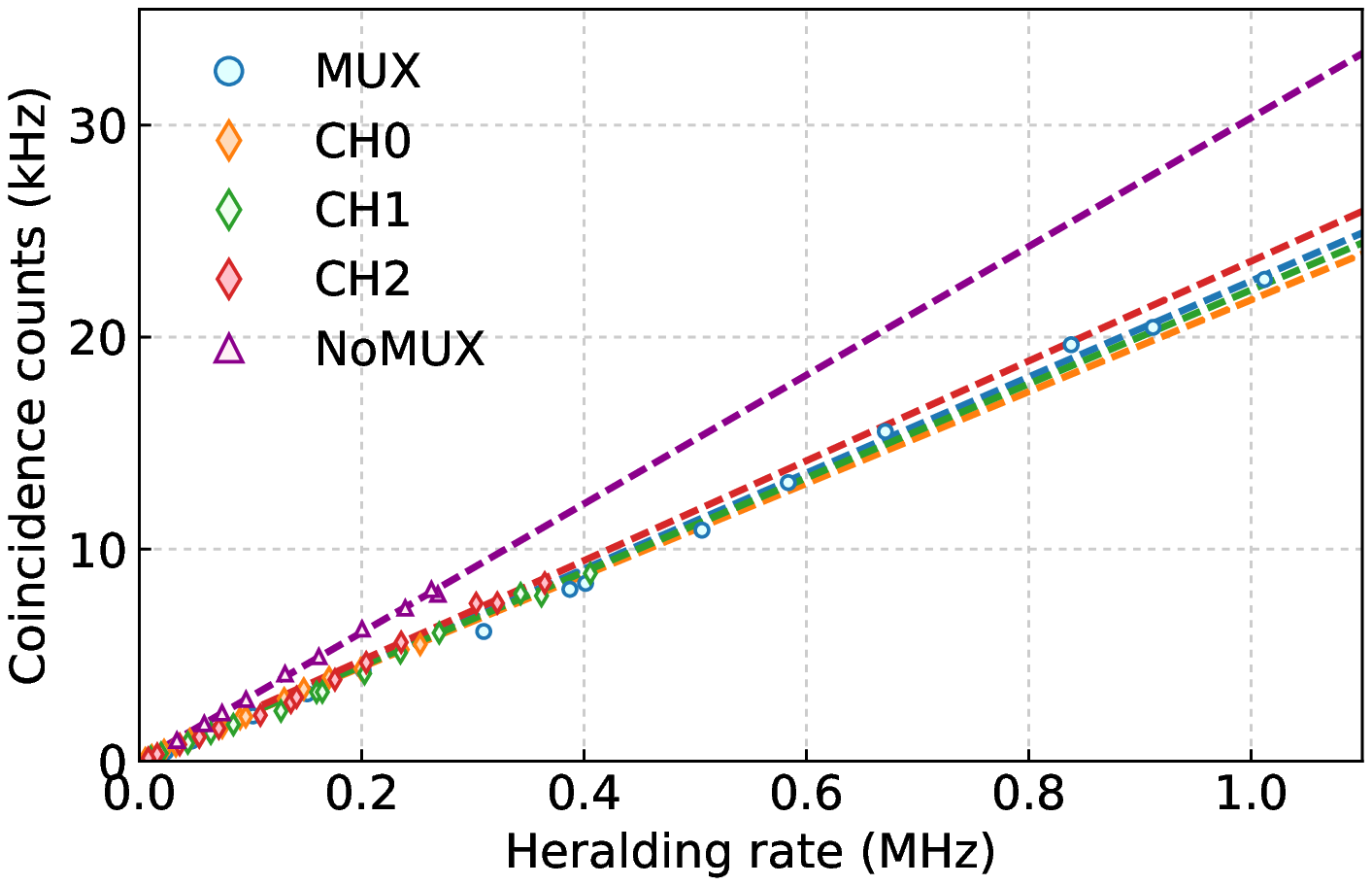}\caption{\label{fig:Characterization-of-heralding}Characterization of heralding
efficiency. The efficiency is given by the slope of the heralding
rate vs heralded photon rate (measured as coincidence counts). We
measure a heralding efficiency of 2.3\% with the multiplexing setup
in place and an efficiency of 3\% without. This corresponds with the
estimated $1.3$ dB transmission loss through the BS-FWM setup measured
using a classical input. }
\end{figure*}

We measure the heralding efficiency on the multiplexed photon arm
by measuring the ratio of the detected coincidences to the heralding
rate, as shown in Fig. \ref{fig:Characterization-of-heralding}. The
raw heralding efficiency is about 2.3\%. Without the multiplexing
setup in place, we measure a heralding efficiency of 3\%, corresponding
to a 1.3 dB loss due to the multiplexing BS-FWM setup. The losses
in the path of the multiplexed photon after collection from the SPDC
source were measured to be: 1.3 dB BS-FWM setup (WDMs and nonlinear
fiber), 1 dB free-space filtering grating, 2.5 dB fiber-coupling after
filtering and 3 dB detection loss. After accounting for detection
loss, we infer a heralding probability of 4.6\%. We estimate about
8 dB losses at collection from the SPDC source, primarily due to mode-mismatch
and transmission loss from filtering optics. 

\section{comparison with previous demonstrations of multiplexed photon sources}

In Table \ref{tab:Comparison-of-the-1}, we compare the the performance
characteristics of our frequency multiplexed source with previous
demonstrations of multiplexing using different schemes and platforms.
Generated rates are calculated after accounting for detector inefficiency.
We note that the maximum switching speed of our system is only limited
by the amplification required for the BS-FWM pumps. In principle,
our system can be operated at a repetition rate of $1/\Delta\nu_{BS}$
where $\Delta\nu_{BS}$ is the acceptance bandwidth of BS-FWM, which
is 100 GHz for our current implementation. 

\section{Frequency Multiplexing scaling}

Here we show how an implementation of frequency multiplexing using
BS-FWM can support large number of frequency channels without drop
in conversion efficiency and achieve a 50\% single-photon heralding
probability with just 10 multiplexed modes. 

We introduce for convenience average frequency $\Delta\Omega=(\omega_{P1}+\omega_{P2})/2-\omega_{ZDW}$
and of the frequency offset $\widetilde{\omega}=\omega_{i}-(\Delta\omega/2+\Delta\Omega+\omega_{ZDW})$,
where $\Delta\omega=\omega_{P1}-\omega_{P2}$ is the separation between
the pumps (see \ref{fig:braggconversion}a). The phase mismatch $k=\beta(\omega_{t})-\beta(\omega_{i})-\beta(\omega_{P1})+\beta(\omega_{P2})$
for a BS-FWM can then be expanded around the zero dispersion frequency
$\omega_{ZDW}$ as follows: 

\begin{multline*}
k = \frac{\beta^{(3)}}{6}\left[\left(\Delta\Omega + \tilde{\omega} - \frac{\Delta\omega}{2} \right)^3  - \left( \Delta\Omega+\tilde{\omega} + \frac{\Delta\omega}{2} \right)^3 \right. \\ \left. - \left(-\Delta\Omega - \frac{\Delta\omega}{2} \right)^3 + \left(-\Delta\Omega + \frac{\Delta\omega}{2} \right)^3\right] + \mathcal{O}(\beta^{(4)})
\end{multline*}
\begin{align}
k = \frac{\beta^{(3)}}{6}\left[3\tilde{\omega}\Delta\omega(\tilde{\omega} + 2\Delta\Omega)\right]
\end{align}

This shows that shows that phase-matching can always be fulfilled
by choosing $\widetilde{\omega}=0$. While the condition $k=0$ is
always satisfied, for large detuning $\Delta\omega$ between the input
and the target frequency $\omega_{t},$ the acceptance bandwidth is
modified and depends on the detuning $\Delta\omega$. The target bandwidth
must match the acceptance bandwidth for optimal conversion. 

\begin{figure*}
\subfloat[]{

\includegraphics[width=0.65\textwidth]{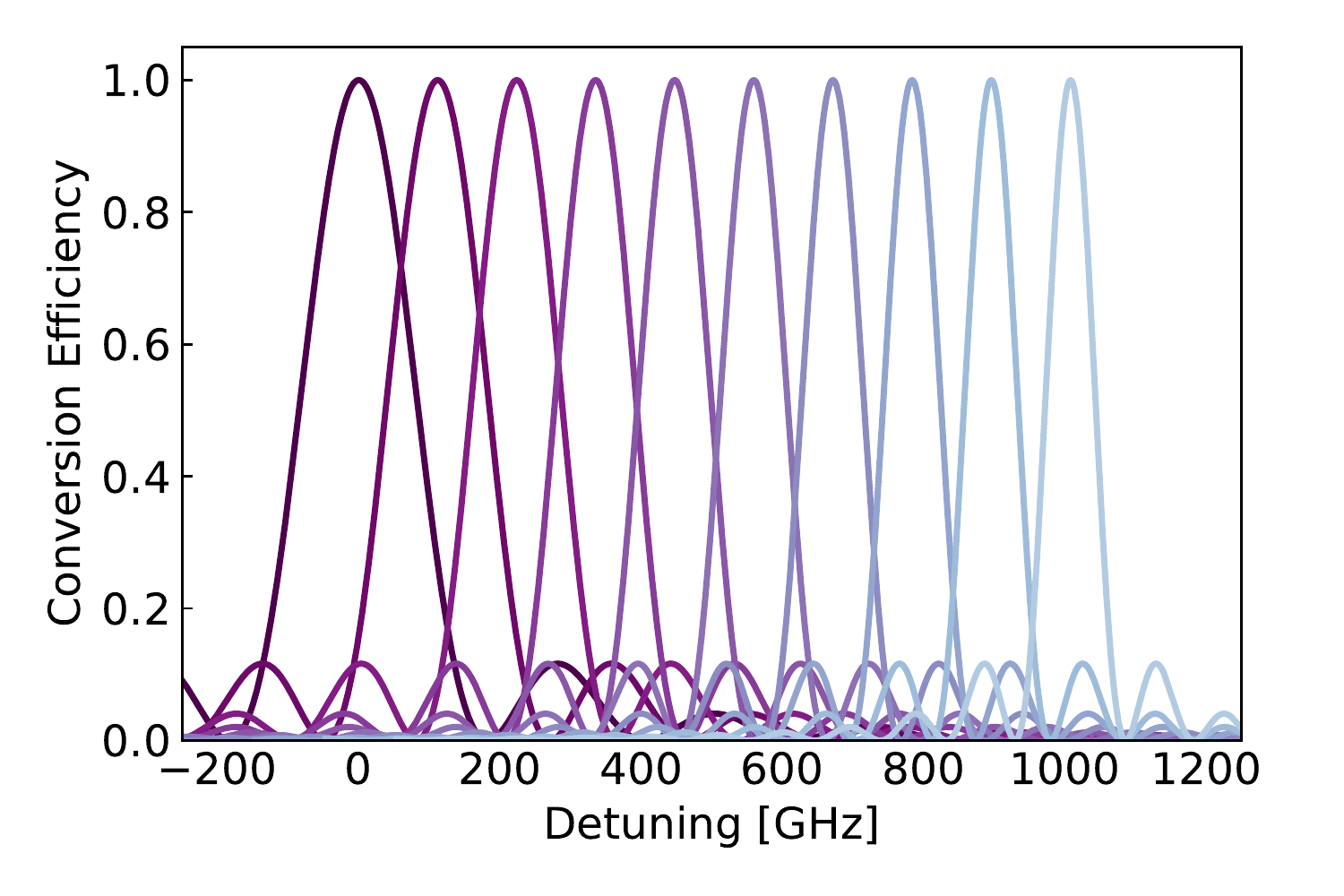}}

\subfloat[]{\includegraphics[width=0.6\textwidth]{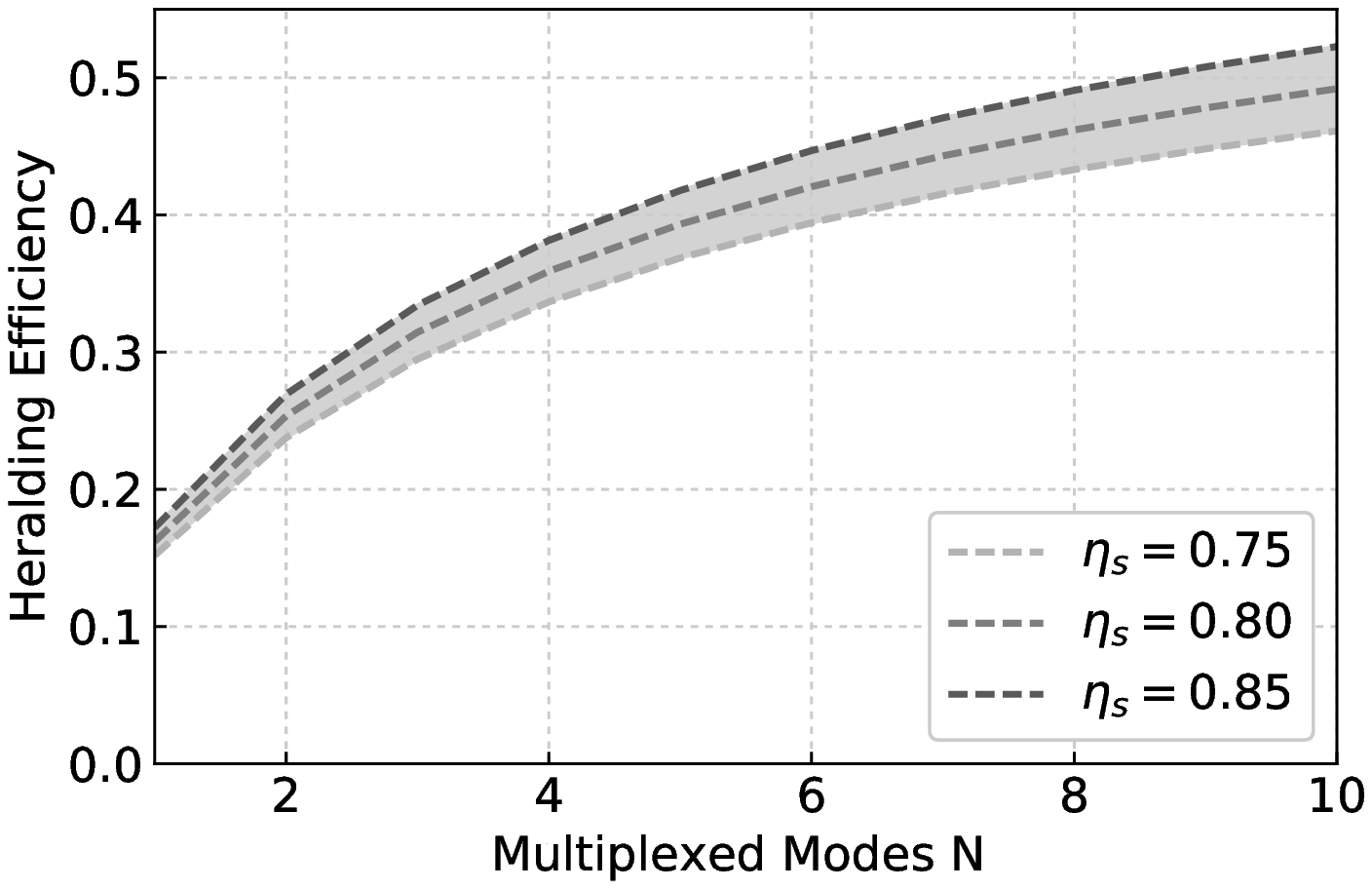}

}

\caption{\label{fig:phasematching-scaling}\textbf{a) }BS-FWM conversion efficiency
for 10 channels separated by 100 GHz. We fix one of the two pumps
while detuning the other such that the frequency separation between
the two pumps matches the frequency separation between the input and
the target. The conversion efficiency is maintained throughout while
the acceptance bandwidth is reduced by a factor of 2 due to the effects
of higher order dispersion. \textbf{b) }Scaling performance of frequency
for 10 frequency modes, assuming a combined detection and fiber-collection
efficiency of 90\%, for varying multiplexing system efficiencies ($\eta_{s}=0.75,0.80,0.85$).
Single-photon heralding efficiencies as high as 50\% can be achieved
with just 10 frequency modes. }

\end{figure*}

In Figure \ref{fig:phasematching-scaling}a we show that we can add
up to 10 additional channels separated by 100 GHz without significant
reduction in the conversion efficiency. The phase-mismatch $k$ is
calculated for the dispersion profile of the nonlinear fiber used
in our experiment, shown in Figure \ref{fig:braggconversion}d. Here,
we assume a fixed target frequency $\omega_{t}$, and we fix one pump
while detuning the other such that the frequency separation between
the pumps equals the frequency separation between the input and target.
The maximum conversion efficiency is maintained for all 10 channels.
The acceptance bandwidth however reduces by a factor of 2, from 160
GHz for the first channel to 70 GHz for the 10th channel. Additional
channels will therefore require stronger filtering. 

Figure \ref{fig:phasematching-scaling}b shows the calculated scaling
performance of the frequency multiplexed source for 10 modes. We assume
a feasible value for detector and fiber-collection efficiency of 90\%.
Heralding efficiencies as high as 50\% can be achieved using just
10 multiplexed modes. 

\end{document}